\documentclass[pre,showpacs,twocolumn]{revtex4}
\usepackage{dcolumn}% Align table columns on decimal point
\usepackage{bm}% bold math
\usepackage[english]{babel}
\usepackage[babel=true]{csquotes}
\usepackage{amssymb}
\usepackage{amsmath}
\usepackage{amsfonts}
\usepackage[]{graphicx}
\usepackage{epstopdf}
\begin{document}
\bibliographystyle{prsty}
\title{Coherence and aberration effects in surface plasmon polariton imaging}
\author{Martin Berthel$^1$, Quanbo Jiang$^1$, Camille Chartrand$^1$, Joel Bellessa$^2$, Serge Huant$^1$, Cyriaque Genet$^3$, Aur\'{e}lien Drezet$^{1\dag}$}
\address{(1)Universit\'e Grenoble Alpes, Institut NEEL, F-38000 Grenoble, France and CNRS, Institut NEEL, F-38042 Grenoble, France}
\address{(2)Institut Lumi\`{e}re Mati\`{e}re, UMR5306 Universit\'{e} Lyon 1-CNRS, Universit\'{e} de Lyon, 69622 Villeurbanne cedex, France}
\address{(3)ISIS, UMR 7006, CNRS-Universit\'e de Strasbourg, 8, all\'ee Monge, 67000 Strasbourg, France}
\email{aurelien.drezet@neel.cnrs.fr} %% email address is required
\begin{abstract}
We study theoretically and experimentally coherent imaging of surface plasmon polaritons using either leakage radiation microscopy through a thin metal film or interference microscopy through a thick metal film. Using a rigorous modal formalism based on scalar Whittaker potentials we develop a systematic analytical and vectorial method adapted to the analysis of coherent imaging involving surface plasmon polaritons. The study includes geometrical aberrations due index mismatch which played an important role in the interpretation of recent experiments using leakage radiation microscopy. We compare our theory with experiments using classical or quantum near-field scanning optical microscopy probes and show that the approach leads to a full interpretation of the  recorded optical images.
\end{abstract}

 \maketitle
\section{Introduction}
\indent Leakage radiation microscopy (LRM)\cite{Hecht} has become in the
last years a versatile and powerful tool to image the
propagation of surface plasmon polaritons (SPPs) \cite{Novotny,Raether} along
a dielectric (air)-metal interface on top of a thin metal film. It
relies on the conversion of such SPPs into leaky light modes in the
dielectric (glass) substrate~\cite{Raether,emrs,emrs2}. These
coherent leaky waves are subsequently recorded with an oil
immersion objective and imaged with a camera. Being a far field
imaging method LRM can be used to obtain information on SPPs in
the direct space as well as in the Fourier (i.e. in-plane
momentum) space~\cite{APL2006}. It has been successfully applied to the
study of several plasmonic devices such as planar lenses and
wave-guides, nano-holes and slits, interferometers and photonic
crystals~\cite{Drezet1,Drezet2,Drezet3,Laluet,Wang,BaudrionOptX2008,LiPRL2013,Zhao,Zhu,hulst},
where it is complementary to near-field scanning optical
methods \cite{Hecht,Bouhelier,HohenauOptX2011} and scanning
tunneling electron microscopy (STM)~\cite{BharadwajPRL2011,tao}. Due to
its high sensitivity LRM constitutes an ideal approach for optical
metrology in the Fourier space of phenomena such as spin-Hall
effect, SPP mode steering, and negative
refraction~\cite{GorodetskiPRL2012,Stein}. Recently, LRM has also
been successfully applied to the new field of quantum plasmonics
involving single leaky SPPs generated by individual photon
emitters~\cite{CucheNL2010,MolletPRB2012}. For practical reasons it was necessary to work with fused silica substrates instead of the usual glass coverslips adapted to  the oil immersion objective. This induced geometrical aberrations which were only qualitatively considered in the analysis of LRM images~\cite{CucheNL2010,MolletPRB2012}.    \\
\indent Despite its long history and its growing importance in the
field of plasmonics, until now only few works focused on the
development of a rigorous theory of LRM.  A general approach taken in refs.~\cite{Drezet1,HohenauOptX2011,BharadwajPRL2011,Marty} is to consider the electromagnetic field  generated by a point-like dipole located in the vicinity of a metal film using the Green dyadic formalism~\cite{Novotny} and the transfer matrix method. However, a good understanding of LRM  requires to take rigorously into account the properties of high numerical aperture objectives needed for imaging the SPP fields. This is particularly clear if we consider that leaky SPPs are emitted at very high angles in a regime where the  paraxial optics approximation breaks down.  This aspect of the problem was not included in previous studies although it can have strong impact of the interpretation of optical imaging as we show in this work. Recently~\cite{PRL}, we presented an analytical theory of SPP imaging by LRM adapted to single dipolar emitter near a metal film without including the geometrical abberations. In ~\cite{PRL} we focused our study more on the physical mechanisms which lie at the heart of leakage radiation that on the optical theory describing the imaging process it self. The aim of this work is therefore  twofold. On the one hand, we give a complete and detailed analytical theory of coherent imaging adapted to LRM and based on transverse electric (TE) and transverse magnetic (TM) scalar Whittaker potentials~\cite{Whittaker}. This approach is well adapted to a clear description of coherent imaging and in particular reduces the  full vectorial description of the field to the knowledge of two scalar functions $\Psi_{\textrm{TE}}$ and $\Psi_{\textrm{TM}}$, the second one being the most important for SPP imaging. We will show that using such scalar potentials strongly simplifies the formalism adapted to LRM since it allows a clear representation of optical observables associated with TE and TM waves.  Moreover, since LRM involves mainly  TM modes    the interpretation and discussion of SPP propagation become easier and reduce to the knowledge of only one complex number $\Psi_{\textrm{TM}}\simeq \Psi_{\textrm{SPP}}$.  On the other hand, in this work, we analyze the  effect
of geometrical aberrations due to defocusing to an index mismatch between the sample and microscope objective on the image coherence. This is a key ingredient for quantum plasmonics using LRM. We will illustrate the theory with experimental images obtained using a near field scanning optical microscope (NSOM). We will consider two kinds of NSOM tips, i.e., either a classical-like aperture NSOM tip, or a quantum-like NV-based NSOM tip.  While the effect of geometrical
aberrations on  confocal microscopy is already well documented the implication of such imperfections  on coherent imaging in particular those involving  propagating SPPs are still unsolved. Due to the long propagation length of SPP mode in the optical regime, i.e., $~10-20\mu$m, the effect cannot be ignored even if the index mismatch is small.  Here we show that the effect is indeed detrimental to coherent imaging and can lead to strongly  distorted images.
\section{Coherent imaging theory}
\begin{figure}[h!t]
\centering
\includegraphics[width=\columnwidth]{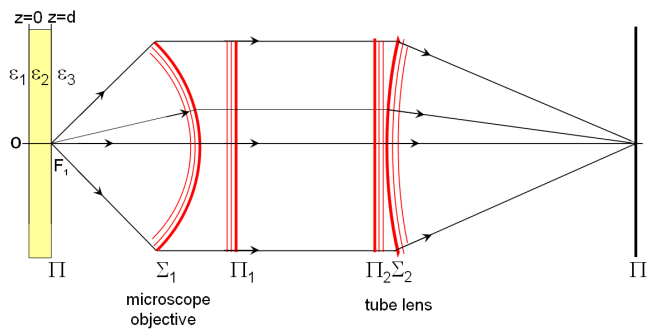}
\caption{(Color online) Sketch of the optical microscope including a high numerical aperture objective. Rays originating from the focus $F_1$ are collimated  along the optical axis after crossing the reference sphere $\Sigma_1$. The  plane $\Pi$ is mapped onto  $\Pi'$. $\Pi_1$ is the back focal plane of the objective and $\Pi_2$, $\Sigma_2$ play for the ocular  or tube lens the same role played by $\Pi_1$, $\Sigma_1$ for the objective. The metal sample of thickness $d$ is located between the plane $z=0$ and $z=d$ (object plane, $\Pi$).}    \label{figure1}
\end{figure}
\indent In order to describe LRM one must consider light propagation
through an oil microscope objective with high numerical aperture
$NA\sim 1.4$ able to collect waves emitted at angles larger
than the critical value $\theta_c\simeq 42^{\circ}$. This imposes to work beyond the usual paraxial approximation.  Indeed, SPPs are emitted at large angle $\theta_{LR}\simeq\arcsin{(n'_{SPP}/n)}$ where $n'_{SPP}$ is the real part of the SPP optical index defined from the SPP in plane wave vector as $k_{SPP}=k_0n_{SPP}(\lambda)$ where $k_0=2\pi/\lambda$ and $\lambda$ is the optical wavelength~\cite{Raether,emrs,emrs2}.  For sufficiently thick films we have $k_{SPP}=k_0\sqrt{(\frac{\epsilon_m}{1+\epsilon_m})}$, with $\epsilon_m$ the complex permittivity of metal, which implies  $n_{SPP}\gtrsim 1$ in the visible. This clearly justifies the use of non paraxial optics to describe LRM. We first fix the geometry
conventions (see Fig.~1). The thin metal film of thickness $d$
supporting propagating SPPs is located on a substratum of optical
index $n\simeq 1.5$ (i.e. glass). The superstratum  is
characterized by a permittivity $\varepsilon_1\simeq 1$ (i.e. air).
The $z$ axis is identified with the optical axis of the microscope
and the different media, i.e., superstratum, metal, substratum are
labeled as media $j=1$, $j=2$, and $j=3$, respectively. Leaky
waves emitted through the metal film propagate in the $+z $
direction through the objective (see Fig.~1). It is customary when
dealing with high numerical aperture lens to define a reference
sphere of radius $f$ associated with the focal length of the
objective~\cite{Wolf1,Wolf2,Torok,Visser}.  This sphere (labeled
$\Sigma_1$) has its center (i.e. the objective focal point) on the
plane $z=d$ which corresponds to the interface between media
$j=2$ and $j=3$. This plane denoted $\Pi$ defines the object plane of the microscope. The wave front located on $\Sigma_1$ evolves into
a planar wave front after propagating through the objective. We
will denote in the following $\Pi_1$ this plane also
identified (as it is usually done) with the back focal plane of
the objective. Clearly, this mathematical treatment of the
objective as a black box does not actually consider the physical
ray propagation in the different lenses which constitute the
objective microscope (including in particular a Weirstrass spherical lens and a meniscus lens). However, this model has shown its efficiency
in the past in particular through study of confocal microscope
setups~\cite{Torok,Visser,Gu}. Finally, light propagates
through a tube lens  (labeled by the planes $\Pi_2$ and
$\Sigma_2$) to reach the image plane $\Pi'$ conjugated with the object plane $\Pi$. However, as we will see, it
is  actually sufficient to model mathematically this lens in the
paraxial regime
where the standard text book description can be used.\\
\indent Having defined the optical setup we now start from
the vectorial Stratton-Chu formulation of Huygens-Fresnel
principle~\cite{Stratton} in order to obtain an integral
representation connecting the electromagnetic field defined at any
point $M$ (i.e. with coordinates $\mathbf{X}=[\mathbf{x},z=d]$) of
the bottom film interface $\Pi$ to the field at point $M_1$ (i.e.
with coordinates $\mathbf{X}_1=[\mathbf{x}_1,z_1]$) on the
reference sphere $\Sigma_1$. Here, we consider the Maxwell
displacement field $\mathbf{D}=\varepsilon_3\mathbf{E}$ with
$\varepsilon_3=n^2$ is the permittivity of the substratum medium
(i.e. glass and oil: $n\simeq 1.5$) and the Stratton-Chu  formula
gives
%\begin{widetext}
\begin{eqnarray}\mathbf{D}_{\Sigma_1}(\mathbf{X}_1)=\int_{(\Pi)}\frac{i\omega\epsilon}{c}\mathbf{\hat{z}}\times\mathbf{B}_\Pi(\mathbf{X}) G_0(R) d^2\mathbf{x}\nonumber\\
+\int_{(\Pi)}[(\mathbf{\hat{z}}\times\mathbf{D}_\Pi(\mathbf{X}))\times
\boldsymbol{\nabla}G_0(R)\nonumber\\
+(\mathbf{\hat{z}}\cdot\mathbf{D}_\Pi(\mathbf{X}))\boldsymbol{\nabla}G_0(R)]
d^2\mathbf{x}
\end{eqnarray}
%\end{widetext}
which is equivalently written, following
Franz~\cite{Franz,Sommerfeld2}, as
\begin{eqnarray}
\mathbf{D}_{\Sigma_1}(\mathbf{X}_1)=\boldsymbol{\nabla}_1\times \int_{(\Pi)}\mathbf{\hat{z}}\times\mathbf{D}_\Pi(\mathbf{X}) G_0(R) d^2\mathbf{x}+\nonumber\\
i\frac{1}{k_0}\boldsymbol{\nabla}_1\times\boldsymbol{\nabla}_1\times
\int_{(\Pi)}\mathbf{\hat{z}}\times\mathbf{B}_\Pi(\mathbf{X}) G_0(R)
d^2\mathbf{x}
\end{eqnarray}
where $G_0(R)=e^{ik_0 nR}/(4\pi R)$ is the usual scalar (Helmholtz)
Green function depending on the distance between points $M$ and
$M_1$: $R=|\mathbf{X}-\mathbf{X}_1|$. Following the Fraunhofer
`far field' approximation one obtains since $R\sim f\gg\lambda$
\begin{eqnarray}
G_0(R)=\frac{e^{ik_0nR}}{4\pi R}\simeq \frac{e^{ik_0nf}}{4\pi
f}e^{-ik_0n\mathbf{x}\cdot\mathbf{x}_1/f}.
\end{eqnarray}
Therefore after some calculations (detailed in appendix A) one
deduces
\begin{eqnarray}
\mathbf{D}_{\Sigma_1}(\mathbf{X}_1)\simeq\frac{2\pi}{if}e^{ik_0nf}k_0n\cos{\theta_1}\tilde{\mathbf{D}}_\Pi[\frac{k_0n\mathbf{x}_1}{f},d]\label{wolf}
\end{eqnarray}
where the polar angle $\theta_1$ is defined by
$f\cos{\theta_1}=z_1-d$ and
$f\sin{\theta_1}=|\mathbf{x}_1|=\varrho_1$. Here
$\tilde{\mathbf{D}}_\Pi[\frac{k_0n\mathbf{x}_1}{f},d]$ is the bidimensional
Fourier transform of the field $\mathbf{D}_\Pi(\mathbf{X})$
defined as $\tilde{\mathbf{D}}_\Pi[\mathbf{k},z]=\int
\frac{d^2\mathbf{x}}{4\pi^2}\mathbf{D}_\Pi(\mathbf{X})e^{-i\mathbf{k}\cdot\mathbf{x}}$
and calculated for
$\mathbf{k}=k_0n\mathbf{x}_1/f=k_0n\varrho_1\boldsymbol{\hat{\varrho}}_1/f$
and for $z=d$. Formula 4 is reminiscent from the work by Wolf
\cite{Wolf1,Wolf2,Novotny} where it is obtained using the
stationary phase approach~\cite{Born} (we will go back to this point later). \begin{figure}[h!t]
\centering
\includegraphics[width=\columnwidth]{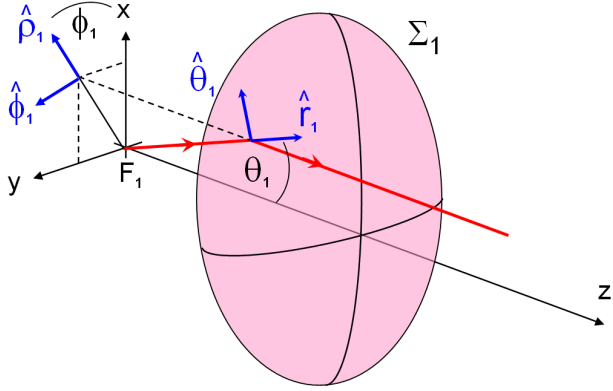}
\caption{(Color online) Sketch illustrating how rays originating from $F_1$ are transformed after crossing the reference sphere $\Sigma_1$. The unit vectors used in the text are indicated. } \label{figure2}
\end{figure}Equivalent calculations, not shown here,
were also done in the reverse case where a collimated light beam is
entering the microscope objective (i.e. in $-z$ direction). In that
case we obtain the results of ref.~\cite{Visser2,Visser2b}. We point
out that $\mathbf{D}_{\Sigma_1}$ is clearly transverse to the sphere
radius joining the focus to $M_1$ as is should be since
$\tilde{\mathbf{D}}_\Pi[\mathbf{k},d]$ is orthogonal to
$\mathbf{k}+\sqrt{k_0^2\varepsilon_3-\mathbf{k}^2}\mathbf{\hat{z}}=k_0n\mathbf{\hat{r}}_1$.\\
In order to describe the propagation through the high NA (aplanatic) objective, we apply the usual `sine'
projection which states that the spherical wave front $\Sigma_1$
evolves into a planar wave front $\Pi_1$ after traveling through
the objective. In this description every conical pencil of light
emerging from the objective focus region  (and intersecting the
reference sphere on a surface element $dS_1$) is therefore
transformed into a cylindrical pencil of light of section
$\cos{\theta_1}dS_1=d^2\mathbf{x}_1$ propagating along a
direction parallel to the optical axis (see Fig.~2). Using the
energy conservation and Poynting theorem one has
\begin{eqnarray}
\frac{t_1}{n^3}|\mathbf{D}_{\Sigma_1}(\mathbf{X}_1)|^2
dS_1=|\mathbf{E}_{\Pi_1}(\mathbf{x}_1,z_{\Pi_1})|^2 d^2\mathbf{x}_1
\end{eqnarray}
where $t_1$ is the transmission of the objective, keeping in mind that for
a plane wave with pulsation $\omega $ propagating along the
direction defined by the unit vector $\mathbf{\hat{k}}$ the time
averaged Poynting vector in a medium with permittivity
$\varepsilon$ and permeability $\mu$ is given by
$\langle\mathbf{S}_{\omega}\rangle=2cRe\{\mathbf{E}_{\omega}\times\mathbf{H}_{\omega}^\ast\}=2c\sqrt{(\varepsilon/\mu)}|\mathbf{D}_{\omega}|^2/\varepsilon^2\mathbf{\hat{k}}$.
The coordinate $z_{\Pi_1}$ of the plane is here supposed to be
associated with the back focal plane of the objective. Taking into
account the vectorial orientation of the (transverse)
electromagnetic field the `sine' condition is written after
separation into TM and TE polarization as:
\begin{eqnarray}
\mathbf{E}_{\Pi_1}(\mathbf{x}_1,z_{\Pi_1})=\frac{T_1}{\sqrt{(n^3\cos{\theta_1})}}\{(\mathbf{D}_{\Sigma_1}(\mathbf{X}_1)\cdot\boldsymbol{\hat{\theta}}_1)\boldsymbol{\hat{\varrho}}_1\nonumber\\
+(\mathbf{D}_{\Sigma_1}(\mathbf{X}_1)\cdot\boldsymbol{\hat{\varphi}}_1)\boldsymbol{\hat{\varphi}}_1\}P_{NA}(\varrho_1).\label{gouzigouzigouzigouzi}
\end{eqnarray}
In this formula we introduce the effective (complex valued) Fresnel
transmission coefficient of the lens $T_1=\sqrt{t_1}e^{i\tau_1}$
which we suppose isotropic and identical for s and p polarizations.
We also include the pupil function of the objective
$P_{NA}(\varrho_1)$ such as $P_{NA}(\varrho_1)=1$ if $\varrho_1\leq
f \sin{\theta_{max}}=f NA/n$ and $P_{NA}(\varrho_1)=0$ otherwise.
For practical application in the Fourier space it is sometimes
better to write Eq.~\ref{gouzigouzigouzigouzi} as
%\begin{widetext}
\begin{eqnarray}
\mathbf{E}_{\Pi_1}(\mathbf{x}_1,z_{\Pi_1})=\frac{2\pi
e^{ik_0nf}}{if}\frac{T_1\sqrt{k_0k_3(k)}}{n}\nonumber\\
\cdot\{(\tilde{\mathbf{D}}_\Pi[\frac{k_0n\mathbf{x}_1}{f},d]\cdot\boldsymbol{\hat{\theta}}_1)\boldsymbol{\hat{\varrho}}_1+(\tilde{\mathbf{D}}_\Pi[\frac{k_0n\mathbf{x}_1}{f},d]\cdot\boldsymbol{\hat{\varphi}}_1)\boldsymbol{\hat{\varphi}}_1\}
\label{backfocal}\end{eqnarray}
%\end{widetext}
where $k_3(k)=\sqrt{k_0^2\varepsilon_3-\mathbf{k}^2}$.\\
\indent The propagation between the objective and the tube lens with focal
length $f'$ can subsequently be treated in the paraxial
approximation. Using Stratton and Chu formalism we obtain
the field $\mathbf{E}_{\Pi_2}(\mathbf{x}_2,z_{\Pi_2})$ in the plane
$\Pi_2$ in front of the tube lens:
\begin{eqnarray}
\mathbf{E}_{\Pi_2}(\mathbf{x}_2,z_{\Pi_2})=\frac{k_0e^{ik_0\Delta}}{2\pi
i\Delta}\int_{\Pi_1}d^2\mathbf{x}_1\mathbf{E}_{\Pi_1}(\mathbf{x}_1,z_{\Pi_1})\nonumber\\
\cdot
e^{-ik_0\frac{\mathbf{x}_1\cdot\mathbf{x}_2}{\Delta}}e^{ik_0\frac{\varrho_1^2}{2\Delta}}e^{ik_0\frac{\varrho_2^2}{2\Delta}}\label{moustic}
\end{eqnarray}
where $\Delta =z_{\Pi_2}-z_{\Pi_1}$ is the the typical tube length of the microscope.\\
The next step is to find the transmitted field through the tube
lens. Using a reasoning similar to the one done for the objective we
get
\begin{eqnarray}
\mathbf{E}_{\Sigma_2}(\mathbf{X}_2)=T_2\sqrt{(\cos{\theta_2})}\{(\mathbf{E}_{\Pi_2}(\mathbf{x}_2,z_{\Pi_2})\cdot\boldsymbol{\hat{\varrho}}_2)\boldsymbol{\hat{\varrho}}_2\nonumber\\
+(\mathbf{E}_{\Pi_2}(\mathbf{x}_2,z_{\Pi_2})\cdot\boldsymbol{\hat{\varphi}}_2)\boldsymbol{\hat{\varphi}}_2\}P_{NA'}(\varrho_2).
\end{eqnarray} where $P_{NA'}(\varrho_2)$ is now defined for the entrance pupil of the lens tube, with the orientation of the z axis $\boldsymbol{\hat{\theta}}_2=\cos{\theta_2}\boldsymbol{\hat{\varrho}}_2+\sin{\theta_2}\boldsymbol{\hat{z}}$ which should be compared to $\boldsymbol{\hat{\theta}}_1=\cos{\theta_1}\boldsymbol{\hat{\varrho}}_1-\sin{\theta_1}\boldsymbol{\hat{z}}$. The final step of the imaging process is to use the Stratton-Chu formula  to describe the electromagnetic field in the region of the image focus. We obtain the so-called vectorial form of the Debye integral  which reads in our case
\begin{eqnarray}
\mathbf{E}_{\Pi'}(\mathbf{x}',z')=\frac{k_0e^{ik_0f'}}{2\pi
if'}\int_{\Sigma_2}\frac{d^2\mathbf{x}_2}{\cos{\theta_2}}\mathbf{E}_{\Sigma_2}(\mathbf{X}_2)\cdot\nonumber\\
e^{-ik_0\frac{\delta z_2\cdot
\delta z'}{f'}}e^{-ik_0\frac{\mathbf{x}_2\cdot\mathbf{x'}}{f'}}e^{ik_0\frac{(\varrho'^2+z'^2)}{2f'}}\label{exitud}
\end{eqnarray}  In this formula $\delta z'$ and $\delta z_2$ are measured relatively to the image focus therefore $\delta z_2/f'=-\sqrt{1-\varrho_2^2/f'^2}\simeq -1+\frac{\varrho_2^2}{2f'^2}$. In the following we will however set  $\delta z'=0$ and work exclusively in the image focal plane. \\
Rigourously speaking Eqs.~4 and 6-10 are sufficient to calculate
the image field formation. Moreover,  for
practical calculation it is possible to assume $\theta_2\ll
\theta_1$. Therefore putting $\theta_2=0$ in Eq.~9 leads to
\begin{eqnarray}
\mathbf{E}_{\Sigma_2}(\mathbf{X}_2)=T_2\mathbf{E}_{\Pi_2}(\mathbf{x}_2,z_{\Pi_2})P_{NA'}(\varrho_2)
\end{eqnarray} and
\begin{eqnarray}
\mathbf{E}_{\Pi'}(\mathbf{x}')=\frac{k_0e^{ik_0f'}e^{ik_0\frac{\varrho'^2}{2f'}}}{2\pi if'}\int_{\Pi_2}d^2\mathbf{x}_2\mathbf{E}_{\Sigma_2}(\mathbf{X}_2)e^{-ik_0\frac{\mathbf{x}_2\cdot\mathbf{x'}}{f'}}.\nonumber\\
\end{eqnarray}
Grouping all of the equations and putting $P_{NA'}(\varrho_2)=1$ for
the range of $\varrho_2$ values considered  lead to the final
expression (see appendix B for details):
\begin{eqnarray}
\mathbf{E}_{\Pi'}(\mathbf{x}')=N\int_{\Pi_1}d^2\mathbf{x}_1\sqrt{(\cos{\theta_1})}P_{NA}(\varrho_1)e^{-ik_0\frac{\mathbf{x}_1\cdot\mathbf{x}'}{f'}}\nonumber\\
\cdot\{(\tilde{\mathbf{D}}_\Pi[\frac{k_0n\mathbf{x}_1}{f},d]\cdot\boldsymbol{\hat{\theta}}_1)\boldsymbol{\hat{\varrho}}_1+(\tilde{\mathbf{D}}_\Pi[\frac{k_0n\mathbf{x}_1}{f},d]\cdot\boldsymbol{\hat{\varphi}}_1)\boldsymbol{\hat{\varphi}}_1\}
\end{eqnarray}
where $N$ is a
constant (see appendix A). The previous formula is equivalently
written as an explicit integral on $\mathbf{k}$ which will be used
in the rest of this work:
\begin{eqnarray}
\mathbf{E}_{\Pi'}(\mathbf{x}')=N'\int_{|\mathbf{k}|\leq k_0NA}d^2\mathbf{k}\sqrt{k_3(k)}e^{-i\mathbf{k}\cdot\frac{\mathbf{x}'}{M}}\nonumber\\
\cdot\{(\tilde{\mathbf{D}}_\Pi[\mathbf{k},d]\cdot\boldsymbol{\hat{\theta}}_1)\boldsymbol{\hat{\varrho}}_1+(\tilde{\mathbf{D}}_\Pi[\mathbf{k},d]\cdot\boldsymbol{\hat{\varphi}}_1)\boldsymbol{\hat{\varphi}}_1\}\label{cornofulgur}
\end{eqnarray} where $N'=\frac{Nf^2}{(k_0n)^{5/2}}$, and  $M=nf'/f\sim 100$ is the magnification of the microscope.\\
\indent All the results discussed insofar are very general and only depend
on the `sine' condition Eq.~\ref{gouzigouzigouzigouzi} valid for
aplanatic microscope objectives. In particular, it should be
observed that only TM modes are geometrically distorted in the
imaging process since the passage from the reference sphere
$\Sigma_1$ to the plane $\Pi_1$ implies a direct modification for
the field. Therefore only a contribution of the TM  field
proportional to $\mathbf{k}$ will survive in the $\Pi_1$ plane. Furthermore, it can be noticed that
all TE waves satisfy already the symmetry requirement for
projection from a spherical  to a plane wave and therefore they  are
not modified by the aplanatic objective lens (up to the transmission
coefficient). This means that in this non paraxial microscopy  phases and
directions of the fields play a critical role when passing from the
Fourier to the image plane $\Pi'$.  Clearly, since SPPs are TM waves emitted at large
$\theta$ angle this means that we cannot ignore the wave front
transformation induced by the objective. This will be the subject of
the next section based on a discussion of Whittaker potentials.
\section{Surface plasmon polariton imaging }
\subsection{The role of the scalar Whittaker  potentials }
\indent We remind that following the pioneer work
by Sommerfeld~\cite{SommerfeldAP1909} the generation process of
leaky SPs (defined in refs.~\cite{burke,burke1,oliner1,oliner2}) by point-like dipoles or current located  in the vicinity of a thin metal layer has been theoretically studied long ago~\cite{Novotny2,Novotny2b,Novotny2c,Novotny,Novotny3}. This
approach has been also recently applied to the context of SP
generation by STM~\cite{Marty,BharadwajPRL2011,tao}. Still, the imaging
procedure itself was essentially ignored partly because it was
observed that classical paraxial optics methods give already a good
quantitative understanding of the propagation~\cite{Drezet4}. This
is nevertheless far from being obvious since leaky SPs are
coherently emitted at a specifical angle~\cite{Raether,emrs}
$\theta_{\textrm{LR}}>\theta_c=\arcsin{[1/n]}\simeq 42
^\circ$  ($\theta_c$  is the critical
angle  of total internal reflection at a glass-air interface).
This corresponds to a regime where paraxial approaches are not
supposed to be true and where the vectorial nature of the
electromagnetic field should neither be neglected. However, it was
recently experimentally suggested that LRM is intrinsically
limited to the imaging of in-plane components of the electric SPP
field~\cite{Wang1,DePeraltaOL2010,HohenauOptX2011} confirming
apparently the intuitive features deduced from a naive paraxial
approximation method.\\
\indent In this context, we remark that the point spread function
of the full microscope for a dipole emitting leaky SPs through the
metal film has been already considered \cite{Sheppard} to describe
the so called surface plasmon coupled emission microscopy (SPCEM)
\cite{Lakowicz,Borejdo,Stefani1,Stefani2}. SPCEM is actually a
particular form of LRM which couples total internal reflection
fluorescence excitation (TIRF) of molecules through a metal film
and LRM in order to enhance the signal-to-noise ratio of standard
TIRF microscopy (i.e. on glass substrate). In refs.
\cite{Stefani1,Stefani2,Kreiter} SPCEM included a scanning
confocal microscope configuration~\cite{Wilson} for which a
precise knowledge of the point spread function mentioned
above~\cite{Sheppard} and involving SPP contributions is required.
For this purpose the approach used in ref.~\cite{Sheppard} is
based on a matrix transfer formalism applied to a plane wave
expansion describing propagation through the metal film.
Importantly, the model includes also a description of the high
NA aplanatic objective in term of a reference
sphere and an integral representation of the electromagnetic field
near a focal point (as given in the general theory by Richards and
Wolf~\cite{Wolf1,Wolf2,Torok,Novotny}). This vectorial formalism
takes into account the transformation of the spherical wave front
emitted by the fluorescent dipoles into a planar wave front
transmitted through the objective (i.e. traveling in the
direction of the tube lens or ocular).\\
\indent For the present purpose we will  however use the scalar potential approach based on the Whittaker expansion proposed in 1904~\cite{Whittaker}
which is, as we will show, more specifically adapted to the analysis of LRM and of 2D coherent imaging.
The first step is the description of the SPP field using a planar modal expansion separating TE and TM components in the three media $j=1,2,3$ corresponding respectively to air, metal, and substrate (i.e. glass or fused silica).  We write for the field in each medium (in a source-free region):
\begin{eqnarray}
\mathbf{D}_j=\boldsymbol{\nabla}\times\boldsymbol{\nabla}\times[\mathbf{\hat{z}}\Psi_{\textrm{TM},j}]+ik_0\varepsilon_j\boldsymbol{\nabla}\times[\mathbf{\hat{z}}\Psi_{\textrm{TE},j}]\nonumber\\
\mathbf{B}_j=\boldsymbol{\nabla}\times\boldsymbol{\nabla}\times[\mathbf{\hat{z}}\Psi_{\textrm{TE},j}]-ik_0\boldsymbol{\nabla}\times[\mathbf{\hat{z}}\Psi_{\textrm{TM},j}]\label{modale2}
\end{eqnarray} with
\begin{eqnarray}
[\boldsymbol{\nabla}^2+k_0^2\varepsilon_j]\Psi_{\textrm{TM,TE},j}=0.
\end{eqnarray}
which indeed shows $\mathbf{B}_{\textrm{TM}}\cdot\mathbf{\hat{z}}=\mathbf{D}_{\textrm{TE}}\cdot\mathbf{\hat{z}}=0$ for fields built only with $\Psi_{\textrm{TM}}$ or $\Psi_{\textrm{TE}}$, respectively.
Now, we have in the glass substrate
\begin{eqnarray}
\mathbf{D}=\boldsymbol{\nabla}\times\boldsymbol{\nabla}\times[\mathbf{\hat{z}}\Psi_{\textrm{TM}}]+ik_0n^2\boldsymbol{\nabla}\times[\mathbf{\hat{z}}\Psi_{\textrm{TE}}]\nonumber\\
=\{\boldsymbol{\nabla}_{||}\partial_z-\boldsymbol{\nabla}_{||}^2\hat{\mathbf{z}}\}\Psi_{\textrm{TM}}]+ik_0n^2\boldsymbol{\nabla}_{||}\times[\mathbf{\hat{z}}\Psi_{\textrm{TE}}]\label{eq15}
\end{eqnarray} We introduce the notation $\mathbf{A}_{||}=\mathbf{A}-(\mathbf{A}\cdot\mathbf{\hat{z}})\mathbf{\hat{z}}$  for any vector (including the Nabla operator) which will be constantly used in this work. The two Whittaker potentials $\Psi_{\textrm{TM,TE}}(\mathbf{X})$ obey the Helmholtz equations
\begin{eqnarray}
[\boldsymbol{\nabla}^2+k_0^2n^2]\Psi_{\textrm{TM,TE}}=0.\label{blob}
\end{eqnarray}
Basic solutions of Eq.~\ref{blob} are given by a Rayleigh-Sommerfeld
expansion
\begin{eqnarray}
\Psi_{\textrm{TM,TE}}(\mathbf{x},z)=\sum_{\pm}\int d^{2}\mathbf{k}\tilde{\Psi}_{\textrm{TM,TE}}^{\pm}[\mathbf{k},z]e^{i\mathbf{k}\cdot\mathbf{x}}\nonumber\\
\textrm{with
}\tilde{\Psi}_{\textrm{TM,TE}}^{\pm}[\mathbf{k},z]=A_{\textrm{TM,TE}}^{\pm}[\mathbf{k}]e^{\pm
ik_3(k)z}.\label{groblob}
\end{eqnarray}
In our problem the 2D Fourier transform gives therefore
\begin{eqnarray}
\tilde{\mathbf{D}}[\mathbf{k},z]
=-\{\mathbf{k}k_3(k)-k^2\hat{\mathbf{z}}\}\tilde{\Psi}^{+}_{\textrm{TM}}[\mathbf{k},z]\nonumber\\
-k_0n^2\mathbf{k}\times\mathbf{\hat{z}}\tilde{\Psi}_{\textrm{TE}}^{+}[\mathbf{k},z]\label{ogourki}
\end{eqnarray}
Remark that we here suppose a field having (real or imaginary)
a wavevector $+k_3(k)$  along the $z$ axis (i.e.
$\tilde{\Psi}_{\textrm{TM,TE}}^{-}[\mathbf{k},z]=0$), in agreement
with causality requirements (Sommerfeld condition). Going back to
Eq.~\ref{ogourki}, we have also
$k_3\mathbf{k}-k^2\hat{\mathbf{z}}=k_0^2n^2\sin{(\theta_1)}\boldsymbol{\hat{\theta}}_1=k_0nk\boldsymbol{\hat{\theta}}_1$
and
$\mathbf{k}\times\hat{\mathbf{z}}=-k_0n\sin{(\theta_1)}\boldsymbol{\hat{\varphi}}_1=-k\boldsymbol{\hat{\varphi}}_1$.
Therefore we obtain an explicit separation of TM and TE waves as:
\begin{eqnarray}
\tilde{\mathbf{D}}_{\textrm{TM}}[\mathbf{k},z]
=-\{\mathbf{k}k_3(k)-k^2\hat{\mathbf{z}}\}\tilde{\Psi}_{\textrm{TM}}[\mathbf{k},z]\nonumber\\=-k_0nk\boldsymbol{\hat{\theta}}_1\tilde{\Psi}_{\textrm{TM}}[\mathbf{k},z]\nonumber\\
\tilde{\mathbf{D}}_{\textrm{TE}}[\mathbf{k},z]
=-k_0n^2\mathbf{k}\times\mathbf{\hat{z}}\tilde{\Psi}_{\textrm{TE}}[\mathbf{k},z]\nonumber\\=+k_0n^2k\boldsymbol{\hat{\varphi}}_1\tilde{\Psi}_{\textrm{TE}}[\mathbf{k},z].\label{farfieldbis}
\end{eqnarray}
\indent We now go back to the derivation of Eq.~\ref{wolf} and consider more specifically the role of Whittaker potentials $\Psi_{\textrm{TM,TE}}(\mathbf{x},z)$ defined by Eq.~\ref{groblob}.  For this purpose we write in the vicinity of the metal  layer for $z\geq d$ (substratum side):
\begin{eqnarray}
\Psi_{\textrm{TM,TE}}(\mathbf{x},z)=\int d^{2}\mathbf{k}\tilde{\Psi}_{\textrm{TM,TE}}[\mathbf{k},d]e^{ik_3(k)(z-d)}e^{i\mathbf{k}\cdot\mathbf{x}}\label{fourier}
\end{eqnarray}
where causality involves only waves propagation along $+z$ in the direction of the observer.
In a recent paper~\cite{PRL} we used the Whittaker potentials to describe the transmitted field generated by a point-like dipole $\boldsymbol{\mu}=\mu_\bot\mathbf{\hat{z}}+\boldsymbol{\mu}_{||}$ located at $z=-h<0$ in the air side. Using the 2D Fourier expansion and writing $\tilde{\Psi}_{\textrm{TM}}[\mathbf{k},d]=\tilde{\Psi}_{\textrm{TM},\bot}[\mathbf{k},d]+\tilde{\Psi}_{\textrm{TM},||}[\mathbf{k},d]$ we get
\begin{eqnarray}\tilde{\Psi}_{\textrm{TM},\bot}[\mathbf{k},d]=\frac{i\mu_\bot}{8\pi^2 k_1}\tilde{T}_{13}^{\textrm{TM}}(k)e^{ik_3 d}e^{ik_1 h}\nonumber\\
\tilde{\Psi}_{\textrm{TM},||}[\mathbf{k},d]=\frac{-i\boldsymbol{\mu}_{||}\cdot\mathbf{k}}{8\pi^2 k^2}\tilde{T}_{13}^{\textrm{TM}}(k)e^{ik_3 d}e^{ik_1 h}\nonumber\\
\tilde{\Psi}_{\textrm{TE}}[\mathbf{k},d]=\frac{ik_0\boldsymbol{\mu}_{||}\cdot(\mathbf{\hat{z}}\times\mathbf{k})}{8\pi^2 k_1k^2}\tilde{T}_{13}^{\textrm{TE}}(k)e^{ik_3 d}e^{ik_1 h}\label{field}\end{eqnarray}
where $k_i(k)=\sqrt{k_0^2\varepsilon_i-\mathbf{k}^2}$ and where we introduce the full transmission Fresnel coefficient  $\tilde{T}_{13}^{\textrm{TM,TE}}(k)$ for both the TM and TE waves, which are defined for the thin metal layer surrounded by air and glass by usual formulas~\cite{PRL}:
\begin{eqnarray}
\tilde{T}_{13}^{\textrm{TM,TE}}(k)=\frac{T_{23}^{\textrm{TM,TE}}T_{12}^{\textrm{TM,TE}}}{1+R_{23}^{\textrm{TM,TE}}R_{12}^{\textrm{TM,TE}}e^{2ik_2d}}e^{i(k_2-k_3)d}
\end{eqnarray} where
\begin{eqnarray}
R_{ij}^{\textrm{TM}}=\frac{k_i/\varepsilon_i-k_j/\varepsilon_j}{k_i/\varepsilon_i+k_j/\varepsilon_j}\nonumber\\
T_{ij}^{\textrm{TM}}=\frac{2k_i/\varepsilon_i}{k_i/\varepsilon_i+k_j/\varepsilon_j}\nonumber\\
R_{ij}^{\textrm{TE}}=\frac{k_i-k_j}{k_i+k_j}\nonumber\\
T_{ij}^{\textrm{TE}}=\frac{2k_i}{k_i+k_j}.
\end{eqnarray}
We emphasize here the importance of boundary conditions at the air metal and metal glass interfaces in deriving these results.  Furthermore, the definition of the electromagnetic fields in term of Whittaker potentials  outlined in Eq.~\ref{modale2} was here given for the bulk medium in absence of current and dipole sources. The complete theory with source terms~\cite{PRL}, shows that for the point-like dipole considered here the free space solution considered in Eq.~\ref{modale2} is rigorous. For the present purpose concerned with fields in the substratum region these subtleties are nonetheless not relevant.\\
\indent  Moreover, as explained in ref.~\cite{PRL}, the radiated far field can be evaluated by using a contour deformation in the complex plane  following a method used by Sommerfeld~\cite{SommerfeldAP1909,Sommerfeld2}. The application of this method to a thin layer system leads however to much more intricate calculations than for the single interface case treated by Sommerfeld due to the presence of several possible branch cuts and poles which should  be clearly identified before   making the analysis.  The resulting field involves a single leaky SPP mode and a lateral wave which is associated with a Goos-H\"{a}nchen effect in transmission~\cite{PRL}. Both contributions however  can be neglected in the far field where  the main contributing term results from a steepest descent calculation along a specified path~\cite{PRL}. We get \begin{eqnarray}
\Psi_{\textrm{TM,TE}}(\mathbf{x},z)\simeq\frac{2\pi k_0n\cos{\theta}}{ir}e^{ik_0 nr}\tilde{\Psi}_{\textrm{TM,TE}}[\mathbf{k},d],\label{farfield}
\end{eqnarray}
where $r,\theta$ are defined like in Section II by the relation: $|\mathbf{x}|=r\sin{(\theta)}$, $z-d=r\cos{(\theta)}$, i.e., $r=\sqrt{(z-d)^2+\mathbf{x}^2}$ and $\theta \in [0,\pi]$. The wave-vector $\mathbf{k}=k_0n\sin{\theta}\boldsymbol{\hat{\varrho}}$ defines the far field angular spectrum of the point-like source.\\
\indent This formula can be alternatively obtained using the Rayleigh scalar formula:  \begin{equation}\Psi_{\textrm{TM,TE}}(\mathbf{x},z)=\int_{(\Pi)}d^2\mathbf{x'}\Psi_{\textrm{TM,TE}}(\mathbf{x'},d)\partial_{z'}G_D(R,R')|_{z'=d}\end{equation} where $G_D(R,R')=\frac{e^{ik_0nR}}{4\pi R}-\frac{e^{ik_0nR'}}{4\pi R'}$ is the Dirichlet Green function ($R=\sqrt{(\mathbf{x}-\mathbf{x'})^2+(z-z')^2}$, $R'=\sqrt{(\mathbf{x}-\mathbf{x'})^2+(z+z'-2d)^2}$) such as $G_D(R,R')=0$ if $z'=d$. Using the
approximation~\cite{Born} $R\simeq r-\sin{\theta}\boldsymbol{\hat{\varrho}}\cdot\mathbf{x'}-\cos{\theta}(z'-d)$,  $R'\simeq r-\sin{\theta}\boldsymbol{\hat{\varrho}}\cdot\mathbf{x'}+\cos{\theta}(z'-d)$ naturally leads to Eq.~\ref{farfield} quite generally (i.e., independently of any hypothesis concerning plasmons and material properties). Despite this alternative derivation we emphasize the importance for the present work  of the methods based on the 2D Fourier expansion, i.e., Eq.~\ref{fourier}, which admits simple generalization when using multilayered systems and when we consider geometrical aberrations (this problem will be considered in the next subsection).\\
\indent Whatever the method used for the derivation of Eq.~\ref{farfield} we can easily obtain the electromagnetic far field as given by Eq.~\ref{wolf}. For this we consider the definitions
$\mathbf{D}_{\textrm{TM}}=\boldsymbol{\nabla}\times\boldsymbol{\nabla}\times[\mathbf{\hat{z}}\Psi_{\textrm{TM}}]$, $\mathbf{D}_{\textrm{TE}}=+ik_0\varepsilon_3\boldsymbol{\nabla}\times[\mathbf{\hat{z}}\Psi_{\textrm{TE}}]$ and use
Eq.~\ref{farfield} with the approximation that terms containing the radial derivative of $e^{ik_0 nr}$ oriented along $\mathbf{\hat{r}}$ dominate since they are inversely proportional to the optical wavelength which is the smaller typical length in the far field (the electromagnetic field is locally equivalent to a plane wave in this regime). We get:
\begin{eqnarray}
\mathbf{D}_{\textrm{TM,TE}}\simeq\frac{2\pi k_0n\cos{\theta}}{ir}e^{ik_0 nr}\tilde{\Psi}_{\textrm{TM,TE}}[\mathbf{k},d]\mathbf{Q}_{\textrm{TM,TE}}
\end{eqnarray} with $\mathbf{Q}_{\textrm{TM}}=-(k_0n)^2\mathbf{\hat{r}}\times(\mathbf{\hat{r}}\times\mathbf{\hat{z}})=-k_0nk\boldsymbol{\hat{\theta}}$ and $\mathbf{Q}_{\textrm{TM}}=-k_0^2n^3\mathbf{\hat{r}}\times\mathbf{\hat{z}}=k_0n^2 k\boldsymbol{\hat{\varphi}}$.  Comparing with Eq.~\ref{farfieldbis} we get $\tilde{\Psi}_{\textrm{TM,TE}}[\mathbf{k},d]\mathbf{Q}_{\textrm{TM,TE}}=\tilde{\mathbf{D}}_{\textrm{TM,TE}}[\mathbf{k},d]$ and therefore we can directly justify Eq.~\ref{wolf} from our definitions of the Whittaker potentials.\\
\indent From this\begin{figure}
\centering
\includegraphics[width=\columnwidth]{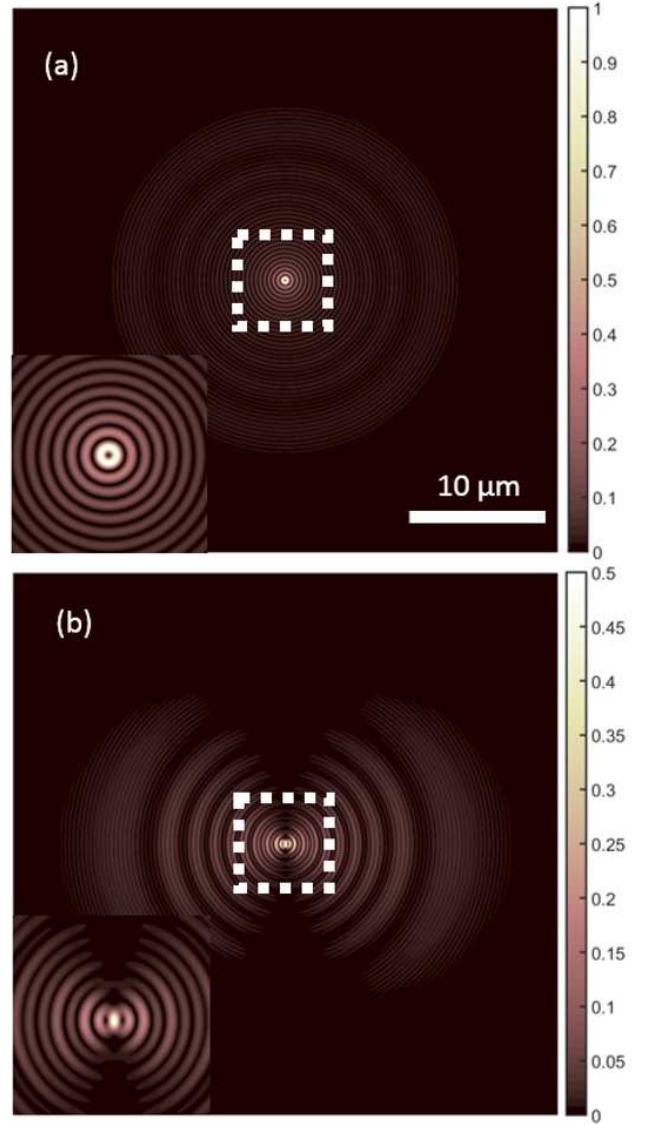}
\caption{(Color online) Simulation of the SPP field imaged in the $\Pi'$ plane for respectively a radiating vertical dipole (a) or horizontal dipole aligned with the x direction (b) and located  on the air side  on top of a 50 nm thick metal film at a distance $h=20$ nm of the air/gold interface. The wavelength is $\lambda=633$ nm and the optical constants for gold are taken from~\cite{JC}. The insets show zooms of the central regions.}   \label{figure3}
\end{figure} modal description all imaging relations obtained in the previous section, in particular Eq.~\ref{backfocal} and \ref{cornofulgur} can be easily translated in terms of Whittaker potentials.
Indeed, at $z=d$ we get
$(\tilde{\mathbf{D}}_\Pi[\mathbf{k},d]\cdot\boldsymbol{\hat{\theta}}_1)\boldsymbol{\hat{\varrho}}_1=-k_0n\mathbf{k}\tilde{\Psi}_{\textrm{TM}}[\mathbf{k},d]=\tilde{\mathbf{D}}_{\textrm{TM},||}[\mathbf{k},d]\frac{k_0n}{k_3(k)}$
while
$(\tilde{\mathbf{D}}_\Pi[\mathbf{k},d]\cdot\boldsymbol{\hat{\varphi}}_1)\boldsymbol{\hat{\varphi}}_1=\tilde{\mathbf{D}}_{\textrm{TE}}[\mathbf{k},d]$. Therefore, regrouping all these expressions we obtain
%\begin{widetext}
\begin{eqnarray}
\mathbf{E}_{\Pi_1}(\mathbf{x}_1,z_{\Pi_1})=\frac{2\pi
e^{ik_0nf}}{if}\frac{T_1\sqrt{k_0k_3(k)}}{n}\nonumber\\
\cdot\{\tilde{\mathbf{D}}_{\textrm{TM},||}[\mathbf{k},d]\frac{k_0n}{k_3(k)}+\tilde{\mathbf{D}}_{\textrm{TE}}[\mathbf{k},d]\}\end{eqnarray}
i.e.
\begin{eqnarray}
\mathbf{E}_{\Pi_1}(\mathbf{x}_1,z_{\Pi_1})=\frac{2\pi
e^{ik_0nf}}{if}\frac{T_1\sqrt{k_0k_3(k)}}{n}\nonumber\\
\cdot\{-k_0n\mathbf{k}\tilde{\Psi}_{\textrm{TM}}[\mathbf{k},d]+k_0n^2k\boldsymbol{\hat{\varphi}}_1\tilde{\Psi}_{\textrm{TE}}[\mathbf{k},d]\}\label{Fourierplane}
\end{eqnarray}
%\end{widetext}
Importantly,  it can be checked that
$|\mathbf{\tilde{D}}_{\textrm{TM},3}[\mathbf{k},d]|^2=|\tilde{\mathbf{D}}_{\textrm{TM},||}[\mathbf{k},d]\frac{k_0n}{k_3(k)}|^2$.
This implies that
\begin{eqnarray}
|\mathbf{E}_{\Pi_1}|^2=\frac{4\pi^2t_1}{f^2n^2}k_0k_3(k)[|\mathbf{\tilde{D}}_{\textrm{TM},3}[\mathbf{k},d]|^2+|\mathbf{\tilde{D}}_{\textrm{TE},3}[\mathbf{k},d]|^2]\nonumber\\
\end{eqnarray}  and therefore that the intensity detected in the back focal plane is proportional to the total Fourier field intensity  for TM and TE waves taken separately. The geometric coefficient $k_3(k)$ shows also that the projection from an infinite plane to an half sphere (neglecting the finite NA) will lead to strong geometrical aberrations at very large angle $\theta$.
Finally, in the image plane Eq.~\ref{cornofulgur} now reads:
\begin{eqnarray}
\mathbf{E}_{\Pi'}(\mathbf{x}')=N'\int_{|\mathbf{k}|\leq
k_0NA}d^2\mathbf{k}\sqrt{k_3(k)}e^{-i\mathbf{k}\cdot\frac{\mathbf{x}'}{M}}
\nonumber\\
\cdot\{\tilde{\mathbf{D}}_{\textrm{TM},||}[\mathbf{k},d]\frac{k_0n}{k_3(k)}+\tilde{\mathbf{D}}_{\textrm{TE}}[\mathbf{k},d]\}
\end{eqnarray}
i.e.
\begin{eqnarray}
\mathbf{E}_{\Pi'}(\mathbf{x}')=N'\int_{|\mathbf{k}|\leq
k_0NA}d^2\mathbf{k}\sqrt{k_3(k)}e^{-i\mathbf{k}\cdot\frac{\mathbf{x}'}{M}}\nonumber\\
\cdot\{-k_0n\mathbf{k}\tilde{\Psi}_{\textrm{TM}}[\mathbf{k},d]+k_0n^2k\boldsymbol{\hat{\varphi}}_1\tilde{\Psi}_{TE}[\mathbf{k},d]\}.\label{cornofulgurbis}
\end{eqnarray} Eqs.~\ref{Fourierplane},\ref{cornofulgurbis} for the imaged field in, respectively, the Fourier and direct planes, which are expressed in terms of Whittaker potentials, are the principal result of this section.\\
\indent In order to illustrate these results we show in Fig.~3 the 2D map in the $\Pi'$ plane of the SP field radiated by a point-like  electric dipole $\boldsymbol{\mu}$ at the optical wavelength $\lambda=633$ nm over a 50 nm gold film (the $z$ coordinate of the dipole is chosen such that the distance $-z=h$ to the air-gold interface equals 20 nm). We focus our interest on a purely vertical dipole (Fig. 3a) and a horizontal dipole aligned in the x direction (Fig. 3b).   First, we observe that in the case of the vertical dipole the intensity map shows a minimum at the center corresponding to the dipole position projected in the $x-y$ plane.
\begin{figure}
\centering
\includegraphics[width=\columnwidth]{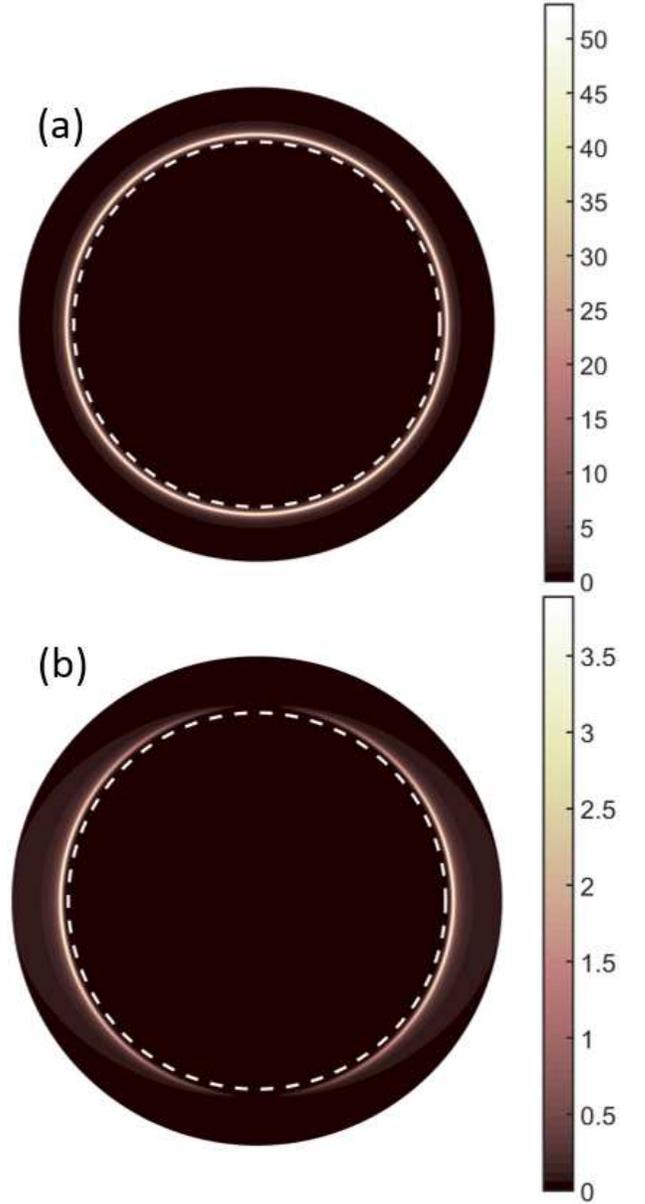}
\caption{(Color online) Fourier space images associated with the real space images of Fig.~3.  (a) is for the vertical dipole and (b) for the horizontal one. The dashed white circle represents unit circle $NA=1$.  } \label{figure4}
\end{figure}
This feature is expected since a vertical dipole cannot radiate energy in the z direction. Furthermore, due to the axial symmetry the electric field is radial in the image plane $\Pi'$  (neglecting the $z$-field component in the paraxial approximation). We have therefore a phase or vortex singularity at the center of the image and the intensity has to vanish in order to preserve the field continuity. In the case of the in-plane dipole the field is not radial but dipolar and we do not observe this vortex anymore.\\
\indent The second important kind of features observed in both Fig. 3a and Fig. 3b is the various radial fringes with small $\Lambda_s=2\pi/(q_1+q_2)$ or large $\Lambda_l=2\pi/(q_1-q_2)$ periodicities. As explained  in Ref.~\cite{HohenauOptX2011} these mainly result from beating between wave components characterized by different  spatial frequencies $q$ associated respectively with the Airy diffraction pattern (dominated by  $q_1= k_0NA$) and with the plasmon wave (with  wave vector $q_2=\textrm{Real}(k_{SP})$). The beating is more pronounced for in-plane dipoles (see Fig.~3b) since the contribution from TE waves is larger in this case compared to the vertical dipole case. This TE contribution, which is not associated with SPPs, is strongly delocalized in the Fourier space and constitutes a background which interfere with the SPP signal and contributes therefore to enhance the fringe visibility  associated with $\Lambda_s$ and $\Lambda_l$. We show in Fig.~4 the Fourier space images obtained for the vertical (Fig. 4a) and horizontal (Fig. 4b) dipoles shown in Fig. 3. The plasmon ring characterized by the wave-vector $q_2$ clearly dominates the images.  The weak contribution associated with TE waves is also visible at large $k$ corresponding to large emission angle in Fig.~4b (for more details on the SPP emission profile see ref.~\cite{PRL}).\\
\begin{figure}
\centering
\includegraphics[width=\columnwidth]{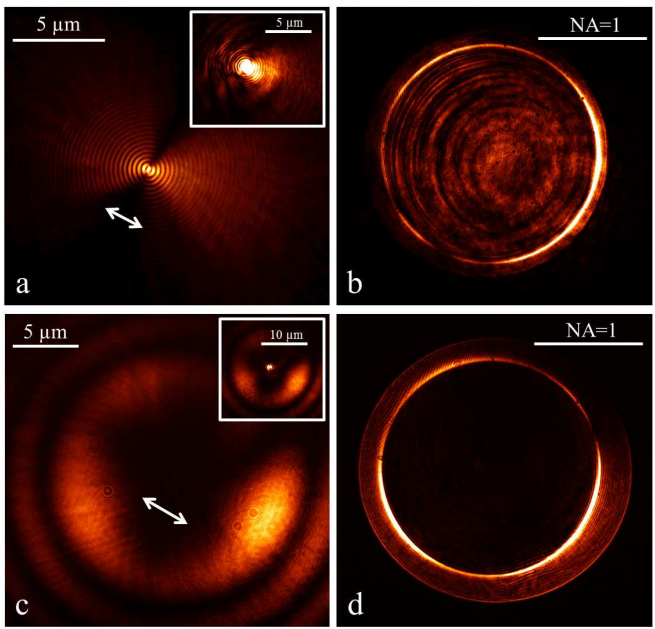}
\caption{(Color online) Experimental LRM images obtained using an aperture NSOM. (a) and (b) correspond to a perfect matching of the glass optical indexes of substrate oil and objective while (c) and (d) are obtained when the glass substrate is replaced by a fused silica substrate with the oil adapted to the new index. (a) and (c) are for the  direct space while (b) and (d) stand for the Fourier space SPP images. (a) is a low $k$-filtered  image showing only the SPP contribution (we included an inset in (a) and (c) to show the unfiltered signal).} \label{figure5}
\end{figure}In order to illustrate the  physics of SPP launching by a point dipole we show in Fig.~5 the experimentally acquired LRM image for a homemade NSOM aperture tip. The fabrication of such a tip is well known and technical details can be found for example in \cite{Novotny,Chevalier}. Here the tip is made of a chemically etched single mode fibre  coated with a 100 nm thick aluminum layer. The 100 nm radius circular aperture located at the apex is the source of light used for near-field optical microscopy. The fiber tip is glued on a quartz tuning-fork and the shear-force coupled to a counter-reaction electronics is used to bring the tip down in the near field and to keep it around 40 nm from the gold surface (our methodology is discussed in refs.~\cite{Brun,Cuche2009,revue}). When light is guided through the fiber down to the aperture ($\lambda=633$ nm), the latter reacts mainly as a pair of electric and magnetic dipoles located in the aperture plane. This pair of orthogonal dipoles behaves as a single equivalent dipole launching SPPs on a flat gold film. Here we show images obtained using a 50 nm thick film evaporated on a glass substrate with optical index $n\simeq 1.52$. The SPP propagation is imaged with a $NA=1.4$ microscope objective using an immersion oil matching exactly the glass index. Fig.~5(a) shows the direct space image obtained using a filter in the back focal plane for masking the low $|\mathbf{k}|$ in-plane momenta corresponding to  $|\mathbf{k}|<k_0$. This opaque mask allows us to filter the directly transmitted light of non plasmonic nature  which is created by the tip. The importance of this effect is case sensitive and depends mainly on the tip aperture diameter and shape. For diameter $<100$ nm the effect is much smaller since the dipolar approximation gets better and better.  Here we also show in the inset the non filtered (saturated) image containing all contributions. The Fourier space LRM image (without the mask) is  shown in Fig.~5(b) for comparison. We see on these images the characteristic features of a in-plane electric dipole launching SPPs on a gold film in good agreement with the theory.  In particular the periodical fringes in the direct space and the ring SPP diameter in the Fourier space agree quantitatively with the model predictions.\\
\indent The case of the vertical dipole has been already studied experimentally using a STM tip~\cite{BharadwajPRL2011,tao} on top of a thin metal film as a SPP launcher. Due to the cylindrical revolution axis the transition dipole is mainly vertical and LRM images  confirm  this finding in agreement with theory. Here, we use instead a Nitrogen vacancy (NV) based NSOM tip to illustrate this vertical dipole feature. The principle of the NV-based NSOM is to attach a diamond nanocrystal (25 nm diameter) containing one or few NV centers whose fluorescence can be excited using a laser ($\lambda=532$ nm) guided through the fiber. The complete protocol of fabrication and use of this system is detailed in a recent review paper~\cite{revue}. Now, \begin{figure}
\centering
\includegraphics[width=\columnwidth]{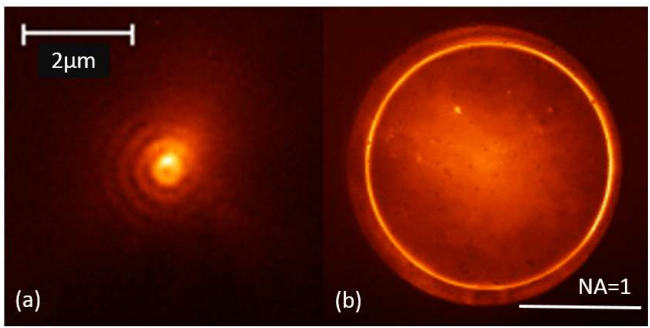}
\caption{(Color online) Experimental LRM images obtained with a NV based NSOM tip over a 50 nm gold film. (a) shows the low $k$-filtered  direct space  SPP image while (b) shows the Fourier space. } \label{figure6}
\end{figure}the point is that if the diamond contains several NVs the probability that a dipole possesses a vertical component is higher. Additionally, in the same conditions (height, film parameters, etc..) a vertical dipole couples better to the SPP modes since the image dipole (in the metal film) is stronger for such a configuration.  Therefore, in general the emission pattern is dominated by the vertical dipole contribution~\cite{PRL}. From a strict theoretical point of view  we can estimate the difference of efficiency by calculating the ratio $\eta=|k/k_1|^2$ at the LRM angle (see Eq.~23). This represents merely the ratio between the SPP field intensity created by respectively a perpendicular and a horizontal dipole.  We obtain $\eta\simeq 13$ which gives a good order of magnitude of the SPP coupling ratio. This is qualitatively  what we show in Fig.~6 where a single nano-diamond containing 5 NVs (the second-order correlation function $g^{(2)}(\tau)$, not shown here, which presents an anti-bunching dip characterizing the quantum nature of our NV source~\cite{revue,Cuche2009,CucheNL2010,MolletPRB2012,Berthel}) has been glued at the apex of a bare  chemically etched tip (without metal coating). The features observed,  and in particular the minimum in the direct plane LRM image, is clearly reminiscent of the vertical dipole calculations shown in Fig.~3(a). We emphasize that in this regime where the signal is extremely weak  we recorded the full broadband fluorescence emission of the  NV center centered in the range $\lambda\simeq 650-750$ nm \cite{MolletPRB2012}.   This lack of temporal coherence clearly affects the low SPP spatial coherence and therefore explains why fringes are hardly visible after few wavelengths in Fig.~6(a). Still, the SPP propagation length estimated from the Fourier space images is for  the film thickness  $d=50$ nm approximately $L_{\textrm{SPP}}\simeq 5-6$ $\mu$m, in agreement with theory (see also ref.~\cite{MolletPRB2012}).    
\subsection{The problem of defocusing and geometrical aberrations}
One of the main issues of this paper is to deal with geometrical aberrations observed with LRM due to a mismatch between the glass substrate optical index $n_g$ and the immersion oil index $n_o$. Indeed, while commercialized immersion microscopes provide their own immersion oil adapted to thin glass cover slips, it can sometimes be useful to use other glass substrates.  This is specially the case in the NV-based NSOM method where the NV fluorescence is excited by a laser light in the $\lambda\sim 515-532$ nm spectral region~\cite{revue,Cuche2009,CucheNL2010,MolletPRB2012}. The problem here is that usual coverslips generate their own fluorescence, in the same spectral band as NVs do, and it becomes impossible to work in the quantum regime, i.e., to build a $g^{(2)}(\tau)$ function presenting an antibunching  $g^{(2)}(0)\simeq 0$ which is the signature of the single-photon emission process~\cite{revue}. To solve this issue we therefore shifted to fused silica coverslips with lower optical index $n'_g\simeq 1.46$ than usual glass $n_g\simeq 1.52$ but  which in turn generates a much lower spurious fluorescence background and are better adapted to microscopy applications with NVs and NSOM.\\
\indent However, while using a fused silica substrate with the good immersion oil index $n_0\simeq n'_g=1.46\equiv n_3$ prohibits image distortion due to geometrical aberration, it is however not possible to eliminate the  mismatch between the oil optical index $n_o=1.46$ and the glass constituting the microscope objective itself $n_g=1.52\equiv n_4$. This mismatch is relatively important and as we will see it is particularly disturbing for LRM due to the high spatial coherence of SPPs. To understand this problem  we will now provide a general study of LRM imaging with defocusing and taking into account an index mismatch between oil and microscope objective.\\
\indent  Consider first the problem of defocusing.   If the glass substrate and oil indexes match the optical index of the objective glass (we will note this common value $n_3$), general Whittaker potentials characterizing the light transmitted through the metal film will be written:
\begin{equation}
\Psi_{\textrm{TM,TE}}(\mathbf{x},z)= \int d^2\mathbf{k}\tilde{\Psi}_{\textrm{TM,TE}}[\mathbf{k},d]e^{ik_3(z-d)}e^{i\mathbf{k}\cdot\mathbf{ x}}
\end{equation}
where $z>d$.  Now if we suppose that the microscope objective is not focussed on the $z=0$ $\Pi$ plane but on the $z=z_F$ $\Pi_F$ plane we can rewrite this expression
\begin{eqnarray}
\Psi_{\textrm{TM,TE}}(\mathbf{x},z)= \int d^2\mathbf{k}\tilde{\Psi}_{\textrm{TM,TE}}[\mathbf{k},d]e^{-ik_3(d-z_F)}\nonumber\\ \cdot e^{ik_3(z-z_F)}e^{i\mathbf{k}\cdot\mathbf{x}}
\nonumber\\
\simeq\frac{2\pi k_3(\theta_3)}{if}e^{ik_0n_3f}\tilde{\Psi}_{\textrm{TM,TE}}[\mathbf{k},d]e^{-ik_3(\theta_3)(d-z_F)}\label{abber}
\end{eqnarray}
In the first line we  simply added and subtracted the phase $k_3z_F$  while in the second line we used a steepest descent method to evaluate this integral  in the far field , i.e. for $z-z_F\gg\lambda$. The angle $\theta_3$ in Eq.~\ref{abber} is the angle made by the $z$ axis  and the emitted ray $FM$ originating from the focus point  $F$ at [$x_F=y_F=0$, $z_F$] and reaching the observation point $M$ located at [$\mathbf{x}$, $z$].
The numerical factor $e^{-ik_3(\theta_3)(d-z_F)}$ in the last expression characterizes completely the geometrical aberration induced on the fields. By including it into the analysis done in the previous section we can describe the effect of defocusing on the image taken in the plane $\Pi'$ which is conjugated with the focal plane $\Pi_F$.\\
\indent However, the problem of interest is \begin{figure}
\centering
\includegraphics[width=\columnwidth]{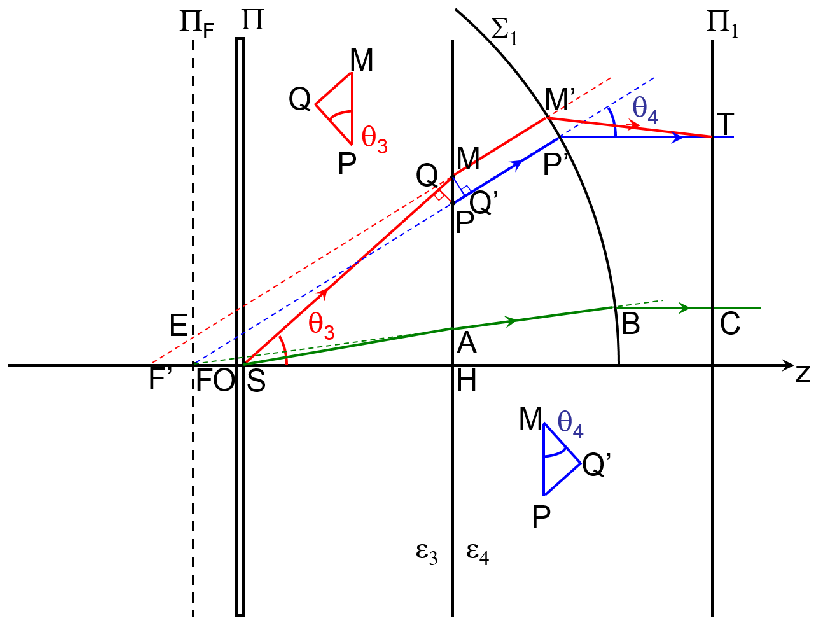}
\caption{(Color online) Sketch of the geometrical rays and interface involved in the spherical aberration modeling (details in the text).  } \label{figure7}
\end{figure}the more general one if there is a an additional  interface at $z_1>>d$ between two media $3$ and $4$ representing the substrate and oil of optical index $n_3$ and the objective microscope glass of index $n_4$, respectively. Writing $D=z_1-d$ the thickness of the substrate plus oil layer (typically $D\simeq 200$ $\mu$m), the Whittaker potentials become:
\begin{eqnarray}
\Psi_{\textrm{TM,TE}}(\mathbf{x},z)\simeq \int d^2\mathbf{k}\tilde{\Psi}_{\textrm{TM,TE}}[\mathbf{k},d]e^{i(k_3-k_4)D}\nonumber\\
\cdot \tilde{T}_{34}^{\textrm{TM,TE}}(k)e^{ik_4(z-d)}e^{i\mathbf{k}\cdot\mathbf{ x}}.
\end{eqnarray}
Here $\tilde{T}_{34}^{\textrm{TM,TE}}(k)$ is the Fresnel transmission coefficient for the $3/4$ interface, and the only approximation is that we will neglect the multiple reflections of light in the layer of thickness $D$. Using the same trick as for Eq.~\ref{abber} we now add and subtract the phase $k_4z_F$ and after using the steepest descent method we obtain in the far field:
\begin{eqnarray}
\Psi_{\textrm{TM,TE}}(\mathbf{x},z)
\simeq\frac{2\pi k_4(\theta_4)}{if}e^{ik_0n_4f}\tilde{\Psi}_{\textrm{TM,TE}}[\mathbf{k},d]\nonumber\\
\cdot e^{-ik_4(\theta_4)(d-z_F)}e^{i(k_3(\theta_3)-k_4(\theta_4))D}.\label{abber2}
\end{eqnarray}
The exponentials on the second lines contain the total additional phase
\begin{eqnarray}
\delta \Phi= (k_3(\theta_3)-k_4(\theta_4))D-k_4(\theta_4)(d-z_F)\label{phasetreuc}
\end{eqnarray}
which  characterizes the geometrical aberration induced in this optical configuration by the index mismatch and defocusing. As before, $\theta_4$ defines the angle between the optical z-axis and  the ray $FP$ originating from the true objective focal point $F$ and reaching the observation point $P$. $\theta_3$  is linked to $\theta_4$ through the Snell-Descartes law $n_4\sin{(\theta_4)}=n_3\sin{(\theta_3)}$.
In order to interpret geometrically $\delta\Phi$  we refer to Fig.~7 and to the following reasoning: First, suppose the source, located at $S$ (at $z=d$), is emitting  a bunch of propagating plane waves traveling through the medium labeled 3. Considering one of these plane waves, the phase accumulated during the straight line propagation from $S$ to the point $M$ allows us to write explicitly the wave phase $\Phi(M)$ at $M$ as $\Phi(M)=k_0n_3 SM=k_0n_3D/\cos{\theta_3}$. Now, using the triangle $QPM$ shown in Fig.~7 the phase at point $P$, which equals the phase at point $Q$, is: $\Phi(P)=\Phi(M)-k_0n_3QM=\Phi(M)-k_0n_3PM\sin{\theta_3}$. Then, since the plane wave is refracted at the 3/4 interface the angle of the wavevector with the $z$-axis is switched from $\theta_3$ to $\theta_4$ through the Snell-Descartes law reminded before. The phase at point $P'$ on the reference sphere  $\Sigma_1$ is $\Phi(P')=\Phi(P)+k_0n_4PP'$. This by definition gives the phase difference $\delta \Phi$ through the relation  $\Phi(P')=\delta\Phi+k_0n_4FP'$, i.e., $\delta\Phi=\Phi(P)-k_0n_4FP$. Taking into account the definition of $\Phi(P)$ given earlier and the geometrical relations $FP=(D+d-z_F)/\cos{\theta_4}$, $PM=D\tan{\theta_3}-(D+d-z_F)D\tan{\theta_4}$ together with the Snell-Descartes law leads directly to Eq.~\ref{phasetreuc}. Here the geometrical reasoning was done for $d-z_F>0$ but similar deductions can be obtained in the opposite case.\\
\indent There are other important features that we can obtain from Fig.~7. Observe indeed that the refraction condition at the 3/4 interface imposes that the wave fronts are in phase at $M$ and $Q'$. Therefore, from the definition of the radius $FP'$ perpendicular to $\Sigma_1$ we know that the waves are also in phase in  $M'$ and $P'$. This is obtained from the relation $n_3QM=n_4Q'P$ which implies $\Phi(P')=k_0[n_3SM-n_3QM+n_4PP']=k_0[n_3SM-n_4Q'P+n_4PP']=k_0[n_3SM+n_4Q'P']=k_0[n_3SM+n_4MM']$, i.e,  $\Phi(P')=\Phi(M)+k_0n_4MM'=\Phi(M')$. Importantly, by definition of the reference sphere $\Sigma_1$ a quasi-plane wave defined as $\Psi(\mathbf{x},z=d)=e^{i\mathbf{k}_0\cdot\mathbf{x}}\Theta(R-|\mathbf{x}|)$, where  $\Theta(u)$ is the unit Heaviside function and $R$ a large radius,  will lead to $\tilde{\Psi}[\mathbf{k},d]=(R^2/4\pi)\frac{2J_1(|\mathbf{k}-\mathbf{k}_0|R)}{|\mathbf{k}-\mathbf{k}_0|R}$. From the results obtained in the first section it means that the far field in the back focal plane will be very much peaked on the wave-vector $\mathbf{k_0}$ when $R$ goes to infinity. The equality $\Phi(P')=\Phi(M')$ associated with such a plane wave was therefore a prerequisite for the self consistency of the calculations. \\
\indent This is not all. From Fig.~7 we see that the ray emerging from $S$ is for an observer located at $M'$ coming virtually from the focus $F'$ which exact location along the z axis is varying with $\theta_3$. However, for low angles $\theta_3$, i.e., in the paraxial regime, $F'$ approaches asymptotically $F$, the true geometrical focus of the objective. From geometrical considerations we have indeed $FF'=D\frac{n_4}{n_3}(\frac{\cos{\theta_4}}{\cos{\theta_3}}-1)$ which cancels in the limit $\theta_3\simeq \frac{n_4}{n_3}\theta_4\rightarrow 0$. Rays in the paraxial regimes are represented in green in Fig.~7 (see the ray $SABC$ in Fig.~7). For an usual NSOM point-like source over a glass substrate, these rays correspond to the light directly transmitted through the sample and strongly contribute to the recorded signal. During operation, when the tip is approximately at point $O$, very close to the surface, the collected signal is optimized only if the objective is translated such that point $F$ is located as sketched on Fig.~7, i.e., not on the interface but on the air side at a distance $FS$ from the surface. From geometrical optics, this can be estimated as: $FS=d-z_F=D(\frac{n_4}{n_3}-1)$. In order to give a order of magnitude for $FS$ we can use the fact that for the appropriate oil and the appropriate glass substrate adapted to the objective ($n_4\simeq 1.52$) \begin{figure}
\centering
\includegraphics[width=8.5cm]{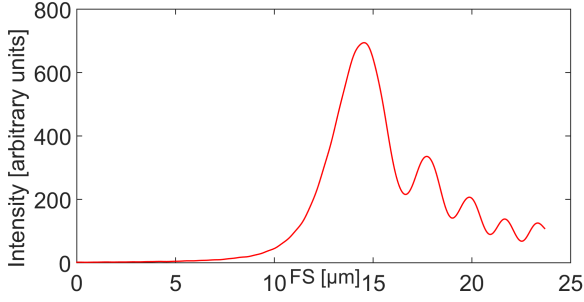}
\caption{(Color online) Recorded LRM intensity of a dipole in the image plane (at the tip image position) as a function of the defocusing $FS$. The maximum is obtained for a distance  $FS=d-z_F\simeq 14.5$ $\mu$m. Here the dipole is along the $x$ direction and $h=20$ nm while the gold film is 50 nm thick. } \label{figure8}
\end{figure}
the focal length $f$ is the sum of the working distance $W_d$, the glass thickness $t_g$ and the objective thickness $L$: $f=W_d+t_g+L$. For the example considered, $W_d=130$ $\mu$m and $D=170$ $\mu$m. Now, when the oil and quartz substrate of optical indexes $n_3=1.46$ are used, the working condition becomes $f=W'_d+t'_g+L+FS$ where $W'_d$ is the new working distance for the quartz (fused silica) substrate of thickness $t'_g\simeq 200$ $\mu$m. Using the definition of $FS$ given earlier with $D=W'_d+t'_g$  leads after elimination of $f$ and $L$ to $W'_d+t'_g=\frac{n_4}{n_3}(W_d+t_g)\simeq 288.15$ $\mu$m and $FS\simeq4\%(W'_d+t'_g)\simeq 11.8$ $\mu$m.
This gives an order of magnitude of the z-displacement with respect to the sample for a optical index mismatch $(n_4-n_3)/n_3\simeq4\%$.\\
However, in order to estimate more rigorously this shift we must go beyond the geometrical optics approximation and use the full field as given by  Eq.~\ref{abber2}. The field in the image plane $\Pi'$ can be numerically obtained by using Eq.~\ref{cornofulgurbis}  and inserting the phase shift  $\delta \Phi$ given by Eq.~\ref{phasetreuc} in the integral.  We show in Fig.~8 the intensity $I(FS)=|\mathbf{E}_{\Pi'}|^2$ recorded in the image plane at the intersection with the $z$ axis when a NSOM tip represented by a point like horizontal dipole aligned with the $x$ axis  is located at $x=y=0$, $z=-20$ nm over the metal-air surface (i.e. at $z=0$). The gold film thickness is chosen  to be $d=50$ nm in order to image SPP leakage radiation.  The maximum of intensity is obtained \begin{figure}
\centering
\includegraphics[width=\columnwidth]{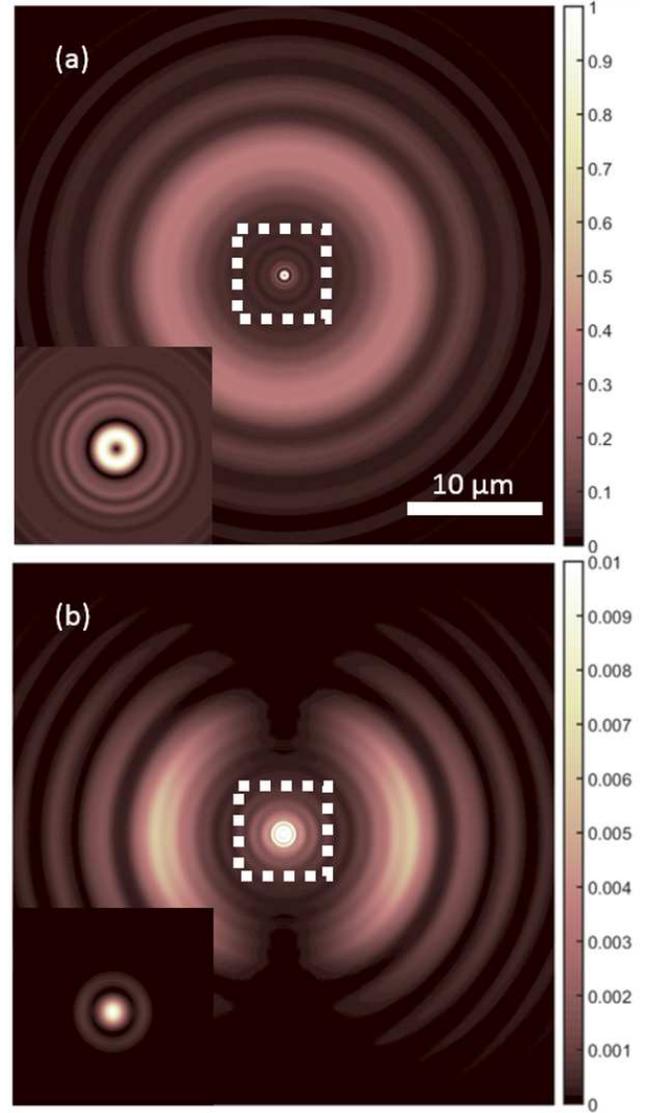}
\caption{Simulation of the LRM aberrated images in the $\Pi'$ plane due to objective/glass-oil index mismatch (see text). (a)  is for the vertical dipole case and (b) for  the horizontal dipole aligned with the x direction (b) and located  on the air side  on top of a 50 nm thick metal film at a distance $h=20$ nm of the air/gold interface (compare with Fig.~3). The wavelength is $\lambda=633$ nm and the optical constants for gold are taken from~\cite{JC}. The insets show unsaturated zooms of the central regions.} \label{figure9}
\end{figure}
for $FS=d-z_F\simeq 14,5$ $\mu$m, which should be compared with the geometrical value obtained before. The discrepancy is attributed to the fact that in the regime considered here SPPs are leaking at large angle, i.e., far beyond the region where the paraxial regime can be applied. \\
\indent It is worth noting that similar analysis was done long ago  to understand confocal fluorescence imaging~\cite{Torok,Visser,Hell}. Here the introduction of a metal layer and the presence of SPPs make the effect more visible. This is better appreciated if we now image in the $\Pi'$ plane the SPP radiation launched by a point like dipole while the microscope objective is positioned at the maximum of intensity defined previously, i.e., for $FS=d-z_F\simeq 14.5$ $\mu$m. We show in Fig.~9 the theoretical images corresponding  to either a vertical dipole (a) or an in-plane dipole along the $x$ direction. The operating conditions are similar to the one shown in Figs.~3,4 ($h=20$ nm, $d=50$ nm). The Fourier space images are not shown since the results are identical to the one observed in Fig.~4. Indeed the common phase factor  $e^{i\delta\Phi}$ does not contribute to the intensity in the Fourier space. The most remarkable features of the images in Figs.~9(a) and 9(b) are the large lobes appearing for distances to the tip around $~10-15$ $\mu$m. The tip itself appears at the center as a well defined `Airy' spot. Note however that for the vertical dipole we observe again the well known donut shaped vortex surrounding the geometrical image that arises from the radial nature of the SPP field radiated by the dipole.\\
\indent In order to compare these simulations with experiments we show in Figs.~5 (c) and 5(d) the recorded NSOM image obtained with an aperture tip over a 50 nm thick gold film (the tip is approximately at $h\simeq 20-30$ nm above the surface). The oil and fused silica index is $n_3=1.46$ while the objective microscope is made of glass with $n_4=1.52$. We clearly observe in the direct space (see Fig.~5(c)) the tip Airy spot together with the large  annular wing surrounding the tip. The Fourier space shows the same general features as already visible in Fig 5 (b) when there is no index mismatch. The experimental images are in good qualitative agreement with Fig.~9(b) and Fig.~4(b) corresponding to a in-plane dipole. Note however that the exact geometry of the tip was unknown and that several multipolar terms can contribute to the observed signal.  In order to understand qualitatively the physical origin of the large wings observed in the image plane we refer to Fig.~10 which should be compared with Fig.~7.
\begin{figure}
\centering
\includegraphics[width=\columnwidth]{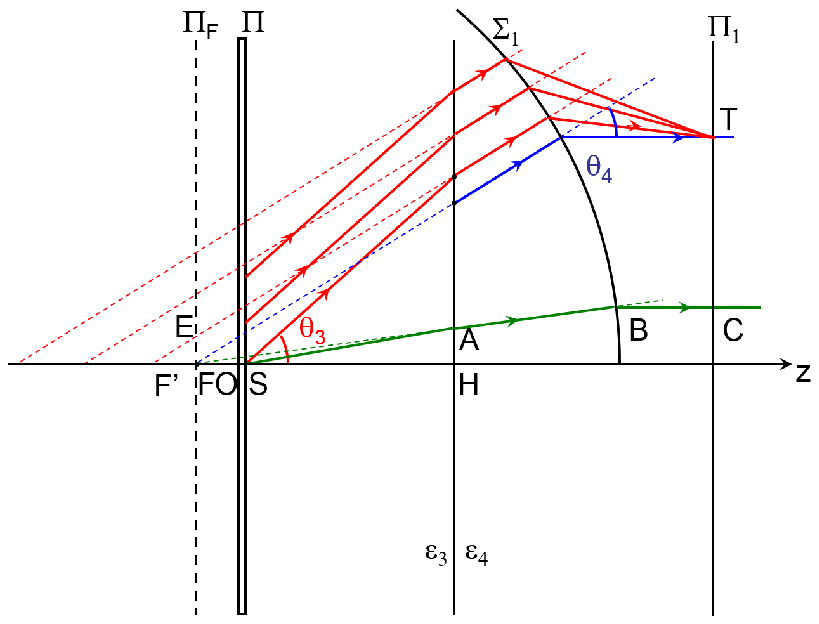}
\caption{(Color online) Sketch of the geometrical  aberrations involved in LRM when an index mismatch modifies the refraction of light at the oil/objective interface (see also Fig.~7) } \label{figure10}
\end{figure}
In the configuration where a point-like dipole near the air metal interface excites  propagating SPPs along the film, leakage radiation will go through the metal layer and produce a conical wave pattern on the glass/oil side (medium 3). This conical wave front presents a shadow zone for diffraction angles below $\theta_{3,LR}\simeq\arcsin{(n_{SPP}/n_3)}$ (see ref.~\cite{PRL} for more details). This means that the signal should ideally be very low for angles below this value. However, due to refraction at the 3/4 interface this conical wave front is distorted and light rays are now virtually coming from virtual planes located before the plane $\Pi_F$ at $z=z_F$. The geometrical plane $\Pi_F$ corresponding to the Airy spot observed at the center of the image plane in Fig.~9, the SPP field will start virtually at a radius $FE$ as visible in Fig.~10. Below this radius there is no SPP field in the image. This radius can be estimated  from the triangle $EFF'$ as $FE=FF'\tan{\theta_{4,LR}}$ where   $\theta_{4,LR}\simeq\arcsin{(n_{SPP}/n_4)}$. Since the leakage angle is close to $45^\circ$ we have $FE\simeq FF'=D\frac{n_4}{n_3}(\frac{\sqrt{1-n_{SPP}^2/n_4^2}}{\sqrt{1-n_{SPP}^2/n_3^2}}-1)\simeq 11.6$ $\mu$m. This is in good qualitative agreement with the observed lobe radius in both the experimental images and the simulations showing that the reasoning picks up the essential parts of the underlying physics.\\
\indent A simple intuitive way to see how this will impact the SPP imaged field is to consider the typical source field $\Psi_{SPP}(\rho)\sim e^{iK_{SPP}\rho}/\sqrt{\rho}$ where $\rho$ is the planar distance to the source on the air/metal interface. This kind of profile characterizes the usual spatial dependency of the radiating SPP field by a dipolar source. Without aberration this field is well imaged on the direct space plane $\Pi'$. Indeed, the fact that the SPP ring  in the Fourier space is well peaked on the wavevector $k'_{SPP}=k_0n'_{SPP}$ with a typical width $k''_{SPP}=1/(2L_{SPP})$ (where the propagation length  $L_{SPP}\simeq 20 -50 $ $\mu$m in the visible for gold and silver) implies that the diffraction/apodization effect associated with the finite $NA\gg n'_{SPP}$ is rather small despite the fringes observed in Fig.~3~\cite{Drezet4}. In particular to a good approximation the image field averaged over a SPP period is equivalent to the real in-plane SPP field at the air/metal interface~\cite{PRL}. In presence of geometrical/spherical aberrations this is of course not true anymore. Due to the radius shift $\rho\rightarrow\rho-FE$ we obtain approximately the following imaged field:
\begin{equation}
\Psi'_{SPP}(\rho)\sim \frac{e^{iK_{SPP}(\rho-FE)}}{\sqrt{(\rho-FE)}}\Theta(\rho-FE)
\end{equation}   where $\Theta(u)$ is the unit Heaviside function. This effect must be taken into account in every LRM images in the $\Pi'$ plane using an optical index mismatch.
\section{Surface plasmon polariton scattering}
In the previous \begin{figure}
\centering
\includegraphics[width=\columnwidth]{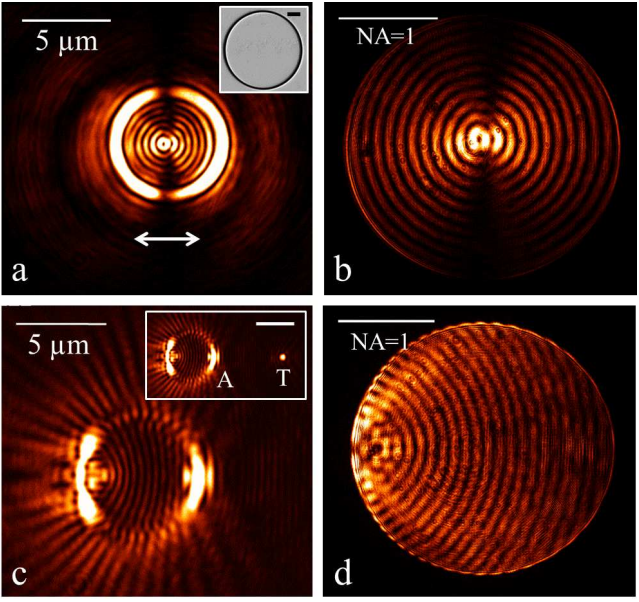}
\caption{(Color online) Experimental images of SPP scattering by a circular slit  (6 $\mu$m diameter) milled on a 200 nm thick gold film. The SPPs are excited by a NSOM tip located at the center of the structure (a) and  (b) or at 12 $\mu$m outside the cavity (c) and (d). (a) and (c) are the direct space images and (b) and (d) the associated Fourier space images. The inset in (a) shows a scanning electron image of the structure (scale bar 1 $\mu$m). The white arrow indicates the tip polarization effective dipole. The tip position $T$ is visible in (c) the distance $AT=9$ $\mu$m (scale bar $5$ $\mu$m).   } \label{figure11}
\end{figure}section we considered the effect of aberrations on SPP imaging through a thin film. Due to the high spatial coherence of SPPs, image distortion can occur in a very large area around the SPP source, i.e., for distances up to typically $\sim $10-20 $\mu$m. However, the problem is not limited to LRM on thin films but is  also impacting the observation of SPPs on thick metal films.\\
\indent In a recent work \cite{JAP} we observed   SPP  induced fringes in the back focal plane of an objective by using a NSOM tip to excite SPPs propagating on a thick film and diffracted by a milled  circular slit acting as a photon coherent source. In this system each point of the circular slit can be described as an in-plane point-like dipole normal to the slit. The coherent sum of all these dipole fields generates  optical fringes in the Fourier plane. This experiment, which is reminiscent of Young's double slit experiments,  allows us to exploit the coherence of SPPs in order to tailor focussed beam such as Bessel modes or polarized vortices~\cite{Wang2,JAP}.  Like for LRM the Fourier plane is insensitive to an index mismatch between the oil and the objective microscope glass.   This is however not true in the direct space plane $\Pi'$.  We compare in Fig.~11 the direct space and Fourier space images for a NSOM aperture located inside a circular cavity made of one slit (width= 150 nm) milled using FIB on a 200 nm thick gold layer on top of a fused silica substrate $n_3=1.46$ (see \cite{JAP} for more details).  We show two situations: either the tip is at the center of the cavity (Figs. 11 (a) and (b)), or located outside at 12 $\mu$m from the center (Figs.~11 (c) and (d)). In both cases we can see physical fringes in the Fourier space (Fig.~11 (b) and (d)), which are clearly reminiscent of the work discussed previously~\cite{Wang2,JAP}. However, if we consider images taken in the direct space, i.e.,  in the $\Pi'$ plane, we can also see optical fringes. These fringes should not be present since the film is opaque and SPPs cannot leak through the metal.\\
They are actually induced by the geometrical aberrations discussed in section 3. Using the formalism developed in  sections 2 and 3 we can justify the existence of these complicated interference patterns. Considering an elementary in-plane point-like electric dipole $\boldsymbol{\mu}$ located on the film we can express the radiated field using a propagator in the fused silica substrate. Indeed, neglecting the geometrical aberrations discussed earlier the displacement field  $\mathbf{D}$ generated by $\boldsymbol{\mu}$ on the surface $\Sigma_1$ is given by
\begin{eqnarray}
\mathbf{D}_{\Sigma_1}(\mathbf{X}_1)=\boldsymbol{\nabla}\times\boldsymbol{\nabla}\times(\frac{\boldsymbol{\mu}e^{ik_0n_3r}}{4\pi r})\nonumber\\
\simeq\frac{k_0^2n_3^2e^{ik_0n_3r}}{4\pi r}[(\boldsymbol{\hat{\theta}}_1\cdot\boldsymbol{\mu})\boldsymbol{\hat{\theta}}_1+(\boldsymbol{\hat{\varphi}}_1\cdot\boldsymbol{\mu})\boldsymbol{\hat{\varphi}}_1]
\end{eqnarray}
From Eq.~40, Eq.~4 and Eqs.~20, 21 we can define the Whittaker potentials associated with with $\boldsymbol{\mu}$ as
\begin{eqnarray}
\tilde{\Psi}_{\textrm{TE}}[\mathbf{k},d]=\frac{ik_0n_3}{8\pi^2 \cos{\theta_3}}(\boldsymbol{\hat{\varphi}}_1\cdot\boldsymbol{\mu})\nonumber\\
\tilde{\Psi}_{\textrm{TM}}[\mathbf{k},d]=\frac{-ik_0n_3}{8\pi^2 }(\boldsymbol{\hat{\varrho}}_1\cdot\boldsymbol{\mu}).
\end{eqnarray}
In this formalism we neglect the boundary conditions \begin{figure}
\centering
\includegraphics[width=\columnwidth]{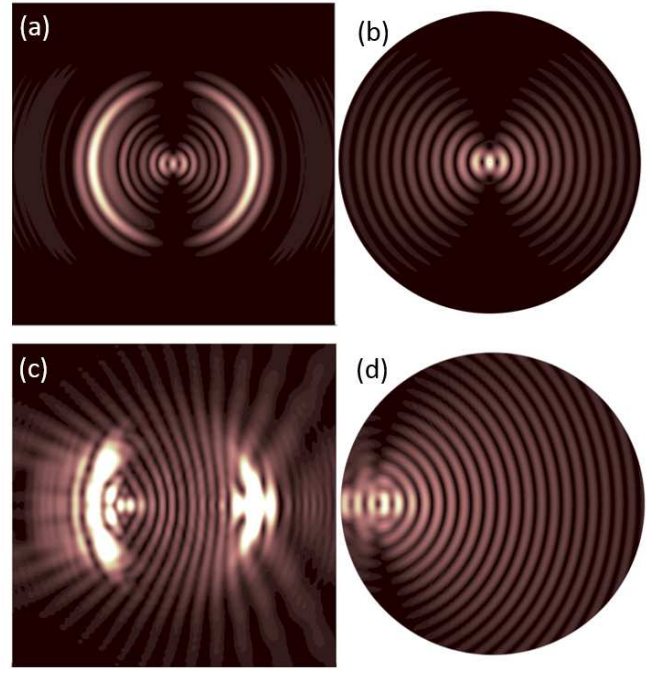}
\caption{(Color online) Simulations of the images obtained in Fig.~11 using the aberration theory developed in this work. (a) and (c) are the direct space images and (b) and (d) the associated Fourier space images. The parameters are the same as for Fig.~11.} \label{figure12}
\end{figure}associated with the fact that the radiation pattern is modified by the presence of the metal. In the case of a perfect metal screen we can show that it is better to describe the emission pattern using magnetic dipoles. However, a quantitative comparison (which will not be shown here) proved that the different ways of describing the field are not easily distinguishable in the regime  considered here. Therefore we will continue to use the simple electric dipole description in the following. Importantly, in order to take into account the geometrical aberrations and the index mismatch we can use the method discussed in section 3 B. In particular from Eqs.~34-38 we can express the new field in presence of aberrations by including the dephasing term $\delta \Phi$ given by  Eq.~38.  The resulting field can be calculated by summing coherently over all dipoles located on the circular slit and excited by the SPP field launched from the NSOM aperture tip (see \cite{JAP} for more details). The simulations corresponding to Fig.~11 are shown in Fig.~12 and show a good agreement between our model and the experiment.

\section{Conclusion}
In this work, we have provided a systematic theory of optical imaging of coherent waves using a high NA objective. Through the use of two scalar Whittaker potentials for TE and TM waves we were able to give transparent expressions for the image fields in the direct and Fourier spaces. Applying this methodology to LRM  we described quantitatively SPP images   generated by point dipoles near  a thin metal film and compared the results with  NSOM measurements provided with either aperture tips or NV based tips. The main experimental issue of this work was to explain and interpret the LRM images obtained when an optical index mismatch  is introduced between the oil and substratum on the one hand and the microscope objective on the other hand.  The theory developed in this work is indeed able to include spherical aberrations generated by this index mismatch, and therefore several questions concerning the interpretation of LRM images obtained in the past are now answered~\cite{CucheNL2010,MolletPRB2012,HohenauOptX2011}.   We also showed that aberrations play a fundamental role for interpreting optical interference images using slits milled on a thick metal film~\cite{JAP}. Again, the theory leads to a clear understanding of the mechanism involved in the experiments.   We expect that the systematic approach developed here together with powerful numerical methods would lead to important progress in the  quantitative interpretation of coherent imaging involving SPPs  in general and in LRM in particular.
\section{Acknowledgments}
This work was supported by Agence Nationale de la Recherche (ANR), France,
through the SINPHONIE and PLACORE grants and the Equipex `Union grant'. The PhD grant of Quanbo Jiang by the
R\'{e}gion Rh\^{o}ne-Alpes is gratefully acknowledged.\\
We thank  Jean-Fran\c{c}ois Motte, from
NANOFAB facility in Neel Institute for the optical tip manufacturing
and FIB milling of the circular slits used in this work.

\section*{Appendix A: The Stratton Chu  and the Wolf formula}
Starting from
\begin{eqnarray}
\mathbf{D}_{\Sigma_1}(\mathbf{X}_1)=\boldsymbol{\nabla}_1\times \int_{(\Pi)}\mathbf{\hat{z}}\times\mathbf{D}_\Pi(\mathbf{X}) G_0(R) d^2\mathbf{x}+\nonumber\\
i\frac{1}{k_0}\boldsymbol{\nabla}_1\times\boldsymbol{\nabla}_1\times
\int_{(\Pi)}\mathbf{\hat{z}}\times\mathbf{B}_\Pi(\mathbf{X}) G_0(R)
d^2\mathbf{x}\label {oriane}
\end{eqnarray} we first evaluate  the integrals $A=\int_{(\Pi)}\mathbf{\hat{z}}\times\mathbf{D}_\Pi(\mathbf{X}) G_0(R) d^2\mathbf{x}$ using the Fraunhofer approximation and we get
\begin{eqnarray}
A=\frac{e^{ik_0nf}}{4\pi
f}\int_{(\Pi)}\mathbf{\hat{z}}\times\mathbf{D}_\Pi(\mathbf{X})e^{-ik_0n\frac{\mathbf{x}_1\cdot\mathbf{x}}{f}}d^2\mathbf{x}\nonumber\\
=\frac{\pi}{f}e^{ik_0nf}\mathbf{\hat{z}}\times\mathbf{\tilde{D}}_{\Pi}[k_0n\mathbf{x}_1/f,d].
\end{eqnarray}
We similarly compute the integral
$\int_{(\Pi)}\mathbf{\hat{z}}\times\mathbf{B}_\Pi(\mathbf{X}) G_0(R)
d^2\mathbf{x}$.  To evaluate Eq.~\ref{oriane}  we use that fact that
the derivative $\mathbf{\hat{r}}_1\partial_1$ along the radial
direction $r_1$ dominates all the other terms, which implies
$\partial_1G_0\simeq ik_0n\mathbf{\hat{r}}_1G_0$.  After regrouping
all the contributions we get:
\begin{eqnarray}
\mathbf{D}_{\Sigma_1}(\mathbf{X}_1)=\frac{i\pi k_0}{f}e^{ik_0nf}[n\mathbf{\hat{r}}_1\times(\mathbf{\hat{z}}\times\mathbf{\tilde{D}}_{\Pi}[k_0n\mathbf{x}_1/f,d])\nonumber\\
-\mathbf{\hat{r}}_1\times(\mathbf{\hat{r}}_1\times(\mathbf{\hat{z}}\times\mathbf{\tilde{B}}_{\Pi}[k_0n\mathbf{x}_1/f,d]))].
\end{eqnarray} Using the fact that we have a transverse plane wave in the Fourier space  we have $n\mathbf{\tilde{B}}=\mathbf{\hat{r}}_1\times\mathbf{\tilde{D}}$ and, therefore,
 \begin{eqnarray}
\mathbf{D}_{\Sigma_1}(\mathbf{X}_1)=\frac{i\pi k_0 n}{f}e^{ik_0nf}\mathbf{\hat{r}}_1\times[\mathbf{\hat{z}}\times\mathbf{\tilde{D}}_{\Pi}[k_0n\mathbf{x}_1/f,d]\nonumber\\
-\mathbf{\hat{r}}_1\times(\mathbf{\hat{z}}\times(\mathbf{\hat{r}}_1\times\mathbf{\tilde{D}}_{\Pi}[k_0n\mathbf{x}_1/f,d]))]\nonumber\\
=\frac{i2\pi k_0
n}{f}e^{ik_0nf}\cos{\theta_1}\mathbf{\tilde{D}}_{\Pi}[k_0n\mathbf{x}_1/f,d]))].
\end{eqnarray}
\section*{Appendix B: The microscope propagator}
Using Eqs.~9,10 together with Eq.~\ref{moustic} leads to
\begin{eqnarray}
\mathbf{E}_{\Pi'}(\mathbf{x}')=-\frac{k_0^2}{4\pi^2 f'\Delta} T_2e^{ik_0(f'+\Delta)}e^{ik_0\frac{\varrho'^2}{2f'}}\nonumber\\
\cdot\int_{\Pi_1}d^2\mathbf{x}_1\mathbf{E}_{\Pi_1}(\mathbf{x}_1,z_{\Pi_1})e^{ik_0\frac{\varrho_1^2}{2\Delta}}\mathcal{I}(\mathbf{x}_1,\mathbf{x}')\label{robert}
\end{eqnarray}
where
\begin{equation}
\mathcal{I}(\mathbf{x}_1,\mathbf{x}')=\int_{(\Pi_2)} d^2\mathbf{x}_2
P_{NA'}(\varrho_2)
e^{-ik_0\mathbf{x_2}\cdot(\frac{\mathbf{x}_1}{\Delta}+\frac{\mathbf{x}'}{f'})}e^{ik_0\frac{\mathbf{x}_2^2}{2\Delta}}.
\end{equation}
The calculation of $\mathcal{I}(\mathbf{x}_1,\mathbf{x}')$ will be explicitly done by supposing $P_{NA'}(\varrho_2)=1$ over the region of interest in the plane $\Pi_2$, which intersects the collimated beam propagating along the $+z$ axis between the objective and the lens tube. This is justified since the radial extension of such a beam in $\Pi_2$ is approximately given by the radius $R_1$ of the exit pupil of the oil immersion objective. Actually, we have $R_1=f\sin{\alpha_1}=NA f'/M $  where $NA=n\sin{\alpha_1}$ is the numerical aperture of the objective. Taking for example $NA=1.4$, $M=100$, and $f'=200$ mm, we deduce $R_1=2.8$ mm  which is in general much smaller than the lens tube radius.\\
Moreover, diffraction of the beam by the exit pupil of the objective leads also to a small angular divergence of the beam $\delta\alpha_1\sim \lambda/R_1$ and therefore to an increase of the beam radius $\delta R_1\simeq\delta\alpha_1\cdot\Delta= \frac{\lambda\Delta}{f'NA}$. If we take, as it is usually the case, $\Delta\simeq f'$, we have $\delta R_1\simeq\frac{\lambda}{NA}$, which at optical wavelength leads to a radius increase of few millimeters. Here, we will altogether neglect the extension $R_1+\delta R_1$  compared  to the radius $R_2$ of the tube lens and we will explicitly integrate over $\mathbf{x}_2$  from $-\infty$ to $+\infty$ along the $x_2$ and $y_2$ direction.\\
For this we use the gaussian integral formula
\begin{equation}
\int^{+\infty}_{-\infty}dx e^{i\alpha x^2}e^{-i\beta
x}=\sqrt{(\frac{i\pi}{\alpha})} e^{-i\frac{\beta^2}{4\alpha}}
\end{equation}
and we get
\begin{equation}
\mathcal{I}(\mathbf{x}_1,\mathbf{x}')=\frac{2\pi i
\Delta}{k_0}e^{-i\frac{k_0\Delta}{2}(\frac{\mathbf{x}_1}{\Delta}+\frac{\mathbf{x}'}{f'})^2}.\label{truc}
\end{equation}
Inserting Eq.~\ref{truc} into Eq.~\ref{robert} leads to
\begin{eqnarray}
\mathbf{E}_{\Pi'}(\mathbf{x}')=\frac{k_0}{2\pi i f'} T_2e^{ik_0(f'+\Delta)}e^{ik_0\frac{\varrho'^2}{2f'}(1-\frac{\Delta}{f'})}\nonumber\\
\cdot\int_{\Pi_1}d^2\mathbf{x}_1\mathbf{E}_{\Pi_1}(\mathbf{x}_1,z_{\Pi_1})e^{-ik_0\mathbf{x}_1\cdot\mathbf{x}'/f'}\label{roberto}
\end{eqnarray} and finally with Eq.~7 we obtain
\begin{equation}
N=-\frac{k_0^2T_1T_2}{ff'\sqrt{n}}e^{ik_0(f'+\Delta+nf)}e^{ik_0\frac{\varrho'^2}{2f'}(1-\Delta/f')}.
\end{equation}
We point out that if we relax the assumption $\delta z'=0$ made in Eq.~\ref{exitud} the main effect  would simply be to add an overall phase factor $\delta \phi=k_0 n\delta z'+k_0 \delta z'^2/2f'$.

\end{document}